\documentclass[12pt,preprint]{emulateapj}
\usepackage{graphicx}

\submitted{Accepted for Publication in The Astronomical Journal: August 14,
2017}
\shorttitle{NGC 6229}
\shortauthors{Johnson et al.}

\newcommand\iso[2]{$^{\rm #1}$#2}

\begin{document}

\title{Light and Heavy Element Abundance Variations in the Outer Halo Globular 
Cluster NGC 6229}

\author{
Christian I. Johnson\altaffilmark{1,2},
Nelson Caldwell\altaffilmark{1}, 
R. Michael Rich\altaffilmark{3}, and
Matthew G. Walker\altaffilmark{4}
}

\altaffiltext{1}{Harvard--Smithsonian Center for Astrophysics, 60 Garden
Street, MS--15, Cambridge, MA 02138, USA; cjohnson@cfa.harvard.edu; 
ncaldwell@cfa.harvard.edu}

\altaffiltext{2}{Clay Fellow}

\altaffiltext{3}{Department of Physics and Astronomy, UCLA, 430 Portola Plaza,
Box 951547, Los Angeles, CA 90095-1547, USA; rmr@astro.ucla.edu}

\altaffiltext{4}{McWilliams Center for Cosmology, Department of Physics,
Carnegie Mellon University, 5000 Forbes Avenue, Pittsburgh, PA 15213, USA;
mgwalker@andrew.cmu.edu}

\begin{abstract}

NGC 6229 is a relatively massive outer halo globular cluster that is primarily
known for exhibiting a peculiar bimodal horizontal branch morphology.  Given
the paucity of spectroscopic data on this cluster, we present a detailed
chemical composition analysis of 11 red giant branch members based on
high resolution (R $\approx$ 38,000), high S/N ($>$ 100) spectra obtained with
the MMT--Hectochelle instrument.  We find the cluster to have a mean 
heliocentric radial velocity of --138.1$_{-1.0}^{+1.0}$ km s$^{\rm -1}$, a 
small dispersion of 3.8$_{-0.7}^{+1.0}$ km s$^{\rm -1}$, and a relatively low
(M/L$_{\rm V}$)$_{\rm \odot}$ = 0.82$_{-0.28}^{+0.49}$.  The cluster is 
moderately metal--poor with $\langle$[Fe/H]$\rangle$ = --1.13 dex and a 
modest dispersion of 0.06 dex.  However, 18$\%$ (2/11) of the stars in our 
sample have strongly enhanced [La,Nd/Fe] ratios that are correlated with 
a small ($\sim$0.05 dex) increase in [Fe/H].  NGC 6229 shares several 
chemical signatures with M 75, NGC 1851, and the intermediate metallicity
populations of $\omega$ Cen, which lead us to conclude that NGC 6229 is a
lower mass iron--complex cluster.  The light elements exhibit the classical
(anti--)correlations that extend up to Si, but the cluster possesses a
large gap in the O--Na plane that separates first and second generation stars.
NGC 6229 also has unusually low [Na,Al/Fe] abundances that are consistent with
an accretion origin.  A comparison with M 54 and other Sagittarius clusters 
suggests that NGC 6229 could also be the remnant core of a former dwarf 
spheroidal galaxy.

\end{abstract}


\section{INTRODUCTION}

Chemical inhomogeneity is a staple of globular cluster formation and evolution,
and the ubiquitous presence of large light element abundance variations has
become a defining characteristic of these systems (e.g., see review by 
\citealt{Gratton12a}).  Although the origins and implications of the light 
element variations remain unsolved problems \citep[e.g.,][]{Bastian15,
Renzini15}, cluster self--enrichment due to pollution from previous generations 
of more massive stars is frequently invoked as a possible explanation 
\citep[e.g.,][]{Decressin07,deMink09,Bastian13,Denissenkov15,D'Antona16}.  In 
this light, it is interesting to note that most clusters do not exhibit similar
abundance variations for the heavier $\alpha$, Fe--peak, and neutron--capture 
elements, and \citet{Carretta09a} have shown that for many cases a cluster's 
intrinsic [Fe/H]\footnote{[A/B] $\equiv$ log(N$_{\rm A}$/N$_{\rm B}$)$_{\rm star}$ -- log(N$_{\rm A}$/N$_{\rm B}$)$_{\sun}$ and log $\epsilon$(A) $\equiv$
log(N$_{\rm A}$/N$_{\rm H}$) + 12.0 for elements A and B.} dispersion is $\la$
12$\%$.  Therefore, most clusters likely completed star formation within the 
first $\sim$100 Myr in order to avoid further pollution by Type Ia supernovae 
([Fe/H] and [$\alpha$/Fe] variations) and/or $\la$ 4 M$_{\rm \odot}$ asymptotic
giant branch (AGB) stars (CNO and s--process variations).

However, a growing number of clusters have been found that possess both light
and heavy element abundance variations.  These ``iron--complex" clusters
share a common set of chemical and physical properties, including: (1) 
intrinsic [Fe/H] spreads that are larger than the measurement errors and 
range from $\sim$12--18$\%$ for clusters such as NGC 1851 and M 75 
\citep[e.g.,][]{Carretta11,Kacharov13} to about a factor of 100 for $\omega$ 
Cen \citep[e.g.,][]{Johnson10,Marino11a}; (2) significant enhancements in 
elements produced by the slow neutron--capture process (s--process) that are 
correlated with [Fe/H]; (3) high cluster masses (M$_{\rm V}$ $\la$ --8); 
(4) very blue and extended horizontal branch (HB) morphologies that frequently 
include significant numbers of extreme HB and blue hook stars; and (5) the 
simultaneous presence of first (O/Mg--rich; Na/Al--poor) and second 
(O/Mg--poor; Na/Al--rich) generation stars in populations with different 
metallicities.  Although iron--complex cluster formation is not yet understood,
the consistent signature of strong s--process enhancements in the more 
metal--rich populations is an indication that these systems experienced 
prolonged ($\ga$ 100 Myr) star formation.  

The unusual chemical and physical characteristics of iron--complex clusters 
suggest that these systems may have different origins than ``normal" 
monometallic clusters.  As summarized in \citet{DaCosta16}, several lines of 
evidence suggest that iron--complex clusters may be the remnant cores
of dwarf spheroidal galaxies that have been accreted and tidally
disrupted by the Milky Way.  For example, several of the most massive 
iron--complex clusters host metal--rich populations that have low 
[$\alpha$/Fe], low [La/Fe], and/or lack the traditional light element abundance
variations associated with globular cluster formation \citep{Carretta10a,
Johnson10,Marino11a,Pancino11,Yong14,Johnson15,Marino15,Johnson17a}, and 
it is possible that these stars represent the original field star populations 
of their parent galaxies.  The two most massive iron--complex clusters 
$\omega$ Cen and M 54 also present strong evidence supporting accretion 
origins.  Simulations indicate $\omega$ Cen could easily survive merging with 
the Galactic disk \citep[e.g.,][]{Bekki03}, and the cluster is known to 
follow a strong retrograde orbit \citep{Dinescu99}.  Similarly, M 54 
resides at the center of the Sagittarius dwarf spheroidal galaxy 
\citep[e.g.,][]{Bellazzini08} and clearly has an extragalactic origin.  Several 
iron--complex clusters also show evidence of diffuse outer envelopes that 
extend beyond their tidal radii \citep{Grillmair95,Olszewski09,Marino14,
Kuzma16}, and these extended halos could represent remnant field 
star populations.

If some fraction of the Galaxy's iron--complex clusters were accreted, then it
is prudent to search for these objects in the outer halo.  Therefore, we 
present an investigation into the chemical composition of the outer halo
(R$_{\rm GC}$ $\sim$ 30 kpc) globular cluster NGC 6229 with the goal of
determining if it is a typical halo cluster, an accreted monometallic cluster,
or a lower mass iron--complex cluster.  \citet{DaCosta16} noted that NGC 6229
is a candidate iron--complex cluster because it has M$_{\rm V}$ $<$ --7.8 and 
exhibits an extended blue HB.  In fact, NGC 6229 shares a similar HB morphology
with the two intermediate metallicity iron--complex clusters M 75 and NGC 1851,
and also NGC 2808 \citep{Borissova97,Borissova99,Catelan98,Catelan02}, which
is known to host at least five populations with different light
element abundances \citep[e.g.,][]{Carretta15,Milone15}.  Some evidence 
suggests that NGC 6229 may be associated with a halo stream and/or the clusters
Pal 4, NGC 7006, and Pyxis \citep{Palma02}, but none of these associations 
have been confirmed \citep[e.g.,][]{Grillmair14}.  Similarly, conflicting 
evidence exists regarding the validity of surface brightness excesses in the 
outer regions of the cluster that may signal the presence of a tidal stream or 
diffuse halo \citep{Sanna12,Carballo14,Bellazzini15}.

\section{OBSERVATIONS AND DATA REDUCTION}

The spectroscopic data for this project were obtained with the MMT 6.5m 
telescope instrumented with the Hectochelle multi--fiber spectrograph
\citep{Szentgyorgyi11} in clear weather on 2016 January 30 and 2016 
April 22--30.  We utilized the ``Cu28" (11 hours), ``CJ26" (7 hours), and
``OB24" (8 hours) filters to obtain high resolution (R $\approx$ 38,000), high 
signal--to--noise (S/N $>$ 100 per reduced pixel) spectra through a series of 
1 hour exposures.  All exposures were binned 2 $\times$ 1 in the spatial and
dispersion directions.  The CU28, CJ26, and OB24 setups provided spectra in the 
5725--5890, 6140--6320, and 6690--6895 \AA\ bands, respectively; however, 
since the fiber output is curved on the detector, the blue/red cut--off
wavelengths are slightly different between fibers located near the center and 
edges of the CCD.  As a result, certain features (e.g., Mg and Al) residing 
near the ends of each order were only measured in some fibers.

\begin{figure*}
\epsscale{1.00}
\plotone{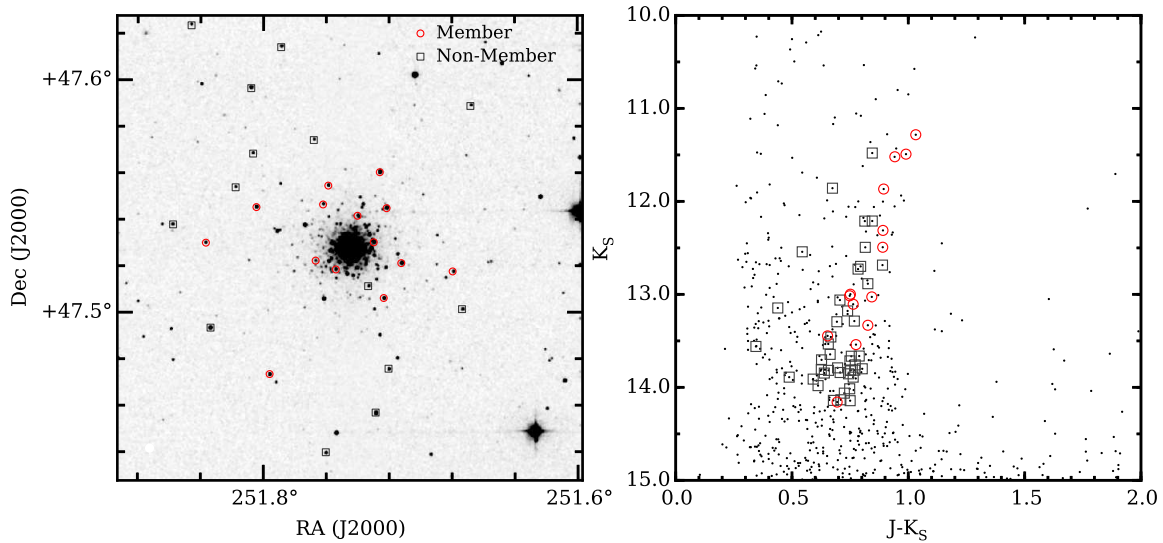}
\caption{Left: the sky coordinates of stars identified as cluster members (open
red circles) and non--members (open grey boxes) are plotted on a 2MASS J--band
image centered near NGC 6229.  The image only spans a radius of 6$\arcmin$,
but additional observations were included out to $\sim$20$\arcmin$; however,
all targets outside 4$\arcmin$ from the cluster center were found to be
non--members.  Right: a 2MASS K$_{\rm S}$ versus J--K$_{\rm S}$
color--magnitude diagram is shown for stars within 20$\arcmin$ of NGC 6229.
Similar to the left panel, the member and non--member stars are designated
with open red circles and grey boxes, respectively.}
\label{f1}
\end{figure*}

Using coordinates and photometry from the Two Micron All Sky Survey (2MASS; 
\citealt{Skrutskie06}), we targeted 57 upper red giant branch (RGB) stars
spanning $\sim$0.5--20$\arcmin$ from the cluster center (see Figure \ref{f1}).
Approximately 30 additional fibers were placed on blank sky regions in order
to obtain simultaneous sky spectra for subtraction.  The targets ranged in 
luminosity from the RGB--tip down to about 1 magnitude brighter than the HB.
Although the membership percentage correlates strongly with radial distance
from the cluster center, the cluster's large distance from the Sun ($\sim$30 
kpc) coupled with Hectochelle's 1.5$\arcsec$ fibers and $\sim$20$\arcsec$ 
positioning constraints prevented placing additional fibers inside 
$\sim$0.5$\arcmin$ of the cluster core with a single configuration.  A
summary of the 2MASS identifiers, J2000 coordinates, and photometry for all
targets is included in Table 1.

A majority of the data reduction was carried out using the dedicated 
Hectochelle pipeline \citep{Caldwell09}, which is maintained by the 
Telescope Data Center at the Smithsonian Astrophysical Observatory.  The 
pipeline handles basic data reduction procedures, such as bias and overscan
corrections, and also performs the fiber tracing, flat--fielding, wavelength
calibration, and spectrum extraction.  Additional processing, such as 
sky subtraction, spectrum co--addition, continuum normalization, and telluric
correction, were carried out independently using standard IRAF\footnote{IRAF is
distributed by the National Optical Astronomy Observatory, which is operated by
the Association of Universities for Research in Astronomy, Inc., under 
cooperative agreement with the National Science Foundation.} tasks.

\section{RADIAL VELOCITIES AND CLUSTER MEMBERSHIP}

Radial velocities were measured for each filter and exposure of every star
with the XCSAO cross--correlation code \citep{Kurtz98}.  The velocities
were measured relative to synthetic RGB spectra of similar temperature and
metallicity that have been smoothed to match Hectochelle's resolution.  A
summary of each star's mean heliocentric radial velocity is provided in 
Table 1 along with the measurement uncertainty, which reflects the standard
deviation of all measurements from each exposure and spectrograph 
configuration for each star.

\begin{figure}
\epsscale{0.80}
\plotone{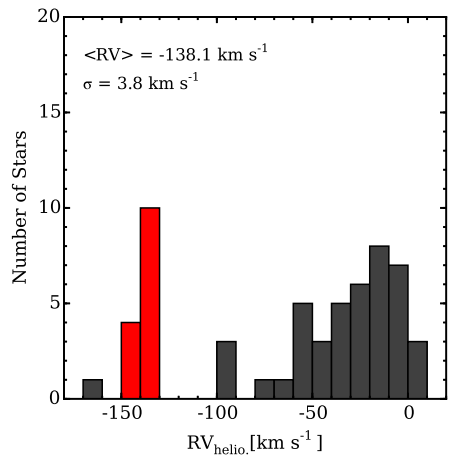}
\caption{A heliocentric radial velocity histogram is shown for all targets
analyzed here.  The filled red and grey histograms indicate stars that are
likely members and non--members, respectively.  The data are sampled into
10 km s$^{\rm -1}$ bins.}
\label{f2}
\end{figure}

Typical measurement uncertainties ranged from about 0.3--0.6 km s$^{\rm -1}$,
but we found one velocity variable star (2MASS 16470843+4744032; non--member)
that varied by $\sim$33 km s$^{\rm -1}$ during the observing period.  A small
number of non--member stars have velocity uncertainties of $\sim$1--2 km 
s$^{\rm -1}$, but these are largely driven by mismatches between the object and 
template spectral types.  Adopting limits of --150 and --130 km s$^{\rm -1}$
(see Figure \ref{f2}), we found 14/57 (25$\%$) stars in our sample to be 
cluster members and all reside within 4$\arcmin$ of the cluster center (see 
Figure \ref{f1}).
   
Following the methodology of \citet{Walker09}, we derive a mean heliocentric
cluster velocity and dispersion of --138.1$_{-1.0}^{+1.0}$ km s$^{\rm -1}$ and 
3.8$_{-0.7}^{+1.0}$ km s$^{\rm -1}$, respectively.  Although few velocity 
measurements are available in the literature for NGC 6229, we note that 
\citet{Pilachowski83} measured a velocity of --139 km s$^{\rm -1}$ from 
one star.  Additionally, a sample of $\sim$10 likely members within 4$\arcmin$
of NGC 6229 from the Apache Point Observatory Galactic Evolution 
Experiment (APOGEE; \citealt{Majewski15}) database yielded a similar velocity
of --138.8 km s$^{\rm -1}$ ($\sigma$ = 4.2 km s$^{\rm -1}$).

Figure \ref{f2} illustrates that the member and non--member stars have clearly
distinct velocity distributions, and therefore the cluster's low velocity
dispersion is unlikely to be affected by field star contamination.  The star
2MASS 16472007+4729363 was found to have a heliocentric velocity of --162.7 km
s$^{\rm -1}$, but it was classified as a non--member because the star's 
spectroscopic temperature did not correlate with its J--K$_{\rm S}$ color.  
Using the \citet{McLaughlin05} structural parameters for NGC 
6229 and the M/L estimator of \citet{Strader09}, we find the cluster to 
have (M/L$_{\rm V}$)$_{\rm \odot}$ = 0.82$_{-0.28}^{+0.49}$.  This is among the
lower M/L ratios of clusters measured with the \citet{Strader09} method,
but is not extreme \citep[e.g.,][]{Strader11}.

\section{DATA ANALYSIS}

\subsection{Model Atmosphere Parameters}

The effective temperatures (T$_{\rm eff}$) listed in Table 2 were determined
by removing trends in plots of log $\epsilon$(\ion{Fe}{1}) versus excitation
potential.  However, since the cluster's line--of--sight reddening is small
(E(B--V) $\approx$ 0.01; \citealt{Zinn80}), we also estimated T$_{\rm eff}$ 
values with dereddened J--K$_{\rm S}$ photometry and the color--temperature 
relation of \citet{GHB09}.  Figure \ref{f3} shows that the spectroscopic and 
photometric T$_{\rm eff}$ estimates are in good agreement with a mean offset, 
in the sense T$_{\rm eff}$ spectroscopic minus T$_{\rm eff}$ photometric, of 
35 K and a dispersion of 116 K.  The star--to--star scatter in Figure \ref{f3} 
is in reasonable agreement with the 94 K T$_{\rm eff}$ uncertainty of the 
color--temperature calibration.  Note that 3/14 member stars were excluded 
from further analysis due to either lower S/N spectra or a significant 
disagreement ($>$ 300 K) between the spectroscopic and photometric 
T$_{\rm eff}$ estimates.

Surface gravities (log(g)) were determined from interpolation within the 
Dartmouth stellar isochrone database \citep{Dotter08} rather than the 
ionization equilibrium of iron.  For each star, we initially adopted the 
log(g) value appropriate for a star with the photometric T$_{\rm eff}$ and 
[Fe/H] = --1.2 dex, and iteratively updated the log(g) value based on new 
estimates of the star's spectroscopic T$_{\rm eff}$ and [Fe/H] abundance.  We 
also assumed [$\alpha$/Fe] = $+$0.2 dex, an age of 12 Gyr \citep{Arellano15}, 
and a canonical He abundance for the isochrone models.

Microturbulence ($\xi$$_{\rm mic.}$) was set by removing trends in plots of 
log $\epsilon$(\ion{Fe}{1}) versus reduced equivalent width (EW).
Lines with log(EW/$\lambda$) $>$ --4.5 were removed in order to avoid 
uncertainties in $\xi$$_{\rm mic.}$ due to line saturation.  Finally, we 
set the model metallicity equal to the derived [Fe/H] abundance, and accounted
for differences between [Fe/H] and [M/H] by adopting the $\alpha$--enhanced
ATLAS9 model atmospheres from \citet{CastelliKurucz04}.

\subsection{Equivalent Width and Spectrum Synthesis Analyses}

The abundances of \ion{Na}{1}, \ion{Al}{1}, \ion{Si}{1}, \ion{Ca}{1}, 
\ion{Cr}{1}, \ion{Fe}{1}, and \ion{Ni}{1} were determined using EWs measured
with the software package outlined in \citet{Johnson14} and the \emph{abfind}
driver of the line analysis code MOOG \citep[2014 version]{Sneden73}.  Isolated
lines were fit with single Gaussian profiles while weakly blended lines were
deblended using multiple Gaussian profiles.  Potential absorption lines were 
identified by visual inspection of a high S/N cluster member spectrum in 
comparison with the Arcturus atlas from \citet{Hinkle00}.  Strongly blended 
lines were flagged and avoided during the analysis.  In an effort to minimize
the impact of departures from local thermodynamic equilibrium and model
atmosphere deficiencies, all abundances were measured relative to Arcturus.
Additionally, log(gf) values were determined by an inverse Arcturus abundance
analysis using the [Fe/H] and [X/Fe] abundance ratios given in 
\citet{Johnson12,Johnson14}.  A summary of the lines selected for measurement, 
along with their lower excitation potentials, log(gf) values, and adopted 
Arcturus and Solar log $\epsilon$(X) abundances, is provided in Table 3.
The final [Fe/H] and [X/Fe] ratios are listed in Tables 4--5.

For \ion{O}{1}, \ion{Mg}{1}, \ion{Cu}{1}, \ion{Zr}{1}, \ion{La}{2}, and 
\ion{Nd}{2}, the abundances were measured using the \emph{synth} spectrum
synthesis module of MOOG.  Since the O abundance of a cool star can be 
affected by its C and N abundances, we simultaneously fit the 6300.3 \AA\ 
[\ion{O}{1}] and nearby CN features using the CN line lists from 
\citet{Sneden14}.  Specifically, the [C/Fe] ratios were set at --0.3 dex for
all stars while we iteratively improved the O and CN synthesis fits by varying 
log $\epsilon$(O) and log $\epsilon$(N).  After determining the CNO abundances,
log $\epsilon$(Mg) was measured using only the bluest line of the 6318 \AA\
\ion{Mg}{1} triplet.  For most stars, the redder triplet lines were too close
to the CCD edge to provide reliable abundances.  The nearby but broad Ca
autoionization feature was fit by adjusting the log $\epsilon$(Ca) abundance
for each star; however, its effects were small ($<$ 0.05 dex) for the stars 
in our sample.  

The \ion{Zr}{1} lines were also fit via spectrum synthesis, but since 
$\sim$90$\%$ of the total Solar Zr abundance is in the even--Z isotopes 
\iso{90}{Zr}, \iso{92}{Zr}, \iso{94}{Zr}, and \iso{96}{Zr} 
\citep[e.g.,][]{Anders89} we did not include any corrections for hyperfine 
broadening.  The Nd abundance is also dominated by even--Z isotopes 
\citep[e.g.,][]{Aoki01,DenHartog03} and did not require additional corrections 
due to isotope wavelength shifts or hyperfine broadening.

The 5782 \AA\ \ion{Cu}{1} line can be significantly affected by a nearby 
diffuse interstellar band (DIB) when the line--of--sight reddening his high.
However, since E(B--V) $\approx$ 0.01 for NGC 6229 we did not detect any
significant DIB features.  The 5782 \AA\ line is relatively strong
and dominated by the two odd--Z nuclei \iso{63}{Cu} (69.17$\%$) and 
\iso{65}{Cu} (30.83$\%$).  Therefore, we derived all Cu abundances using the
hyperfine line list from \citet{Cunha02}.  Similarly, the La abundance is 
dominated by the \iso{139}{La} isotope, which frequently exhibits strong 
hyperfine broadening, and we adopted the line lists of \citet{Lawler01} for
all log $\epsilon$(La) determinations.  A summary of the selected lines and 
[X/Fe] ratios for all elements measured via spectrum synthesis is provided in 
Tables 3--5.

\subsection{Abundance Uncertainties}

\begin{figure}
\epsscale{0.85}
\plotone{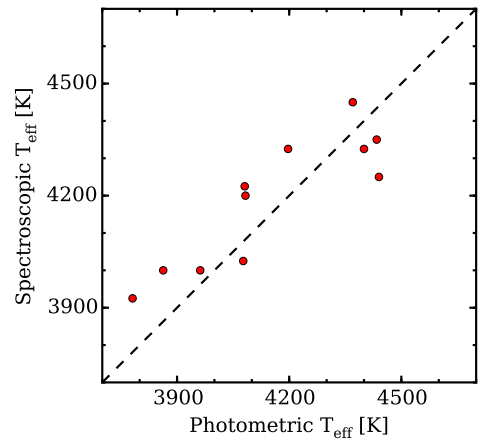}
\caption{Spectroscopic T$_{\rm eff}$ values for stars in NGC 6229 are compared
against photometric T$_{\rm eff}$ estimates derived from dereddened
J--K$_{\rm S}$ photometry and the \citet{GHB09} color--temperature calibration.
The dashed line indicates perfect agreement.}
\label{f3}
\end{figure}

The internal abundance uncertainties are dominated by errors in the model
atmosphere parameter determinations, EW or spectrum synthesis fitting, and 
adopted log(gf) values.  For the model atmosphere values, we have adopted 
typical uncertainties of 100 K in T$_{\rm eff}$, 0.15 (cgs) in log(g), 0.1 dex
in [M/H], and 0.15 km s$^{\rm -1}$ in $\xi$$_{\rm mic.}$.  The temperature
uncertainty is derived from Figure \ref{f3}, which shows that the dispersion
when comparing photometric and spectroscopic T$_{\rm eff}$ estimates is 
approximately 100 K (see also Section 4.1).  By extension, changing 
T$_{\rm eff}$ $\pm$ 100 K also modifies log(g) by $\sim$ 0.15 (cgs) in the 
stellar isochrone set used here (see Section 4.1).  The adopted metallicity
uncertainty of 0.1 dex matches the typical line--to--line [Fe/H] dispersion,
and the $\xi$$_{\rm mic.}$ uncertainty is derived from an examination of the
scatter in plots of log $\epsilon$(\ion{Fe}{1}) versus reduced EW.

Given the high S/N ($>$ 100 per reduced pixel) of the co--added spectra, the 
fitting errors for both EW and synthesis analyses are small ($<$ 0.05
dex).  Additionally, we assume that log(gf) uncertainties are mostly included
in each element's line--to--line abundance scatter, and that model atmosphere
deficiencies are largely offset by keeping our analysis relative to the 
metal--poor giant Arcturus.  Therefore, the abundance uncertainty values given
in Tables 2, 4, and 5 include the line--to--line scatter of each element 
divided by $\sqrt{N}$ and added in quadrature with the cumulative effects of 
changing each model atmosphere parameter by the amount specified above.  For
elements other than iron, the [X/Fe] uncertainties include any correlated 
variations between log $\epsilon$(X) and log $\epsilon$(Fe).

\begin{figure}
\epsscale{0.80}
\plotone{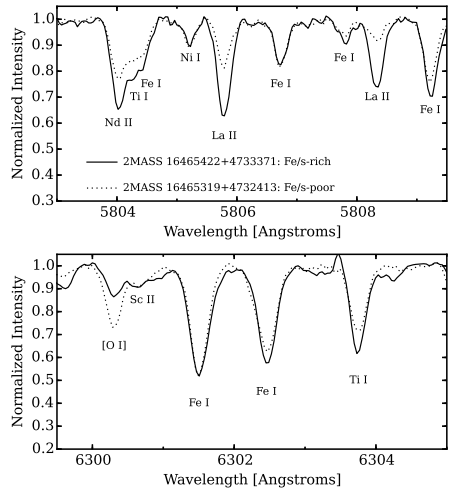}
\caption{Sample spectra are shown illustrating the line strength variations
between two stars with similar temperatures and gravities but very different
light and heavy element abundances.  The O--poor/La--rich star (solid black
lines) may be $\sim$0.05--0.10 dex more metal--rich than its comparatively
O--rich/La--poor counterpart (dotted black lines).}
\label{f4}
\end{figure}

We did not find any significant trends between the derived abundance ratios
and model atmosphere parameters.  However, the three stars with [Fe/H] $>$ 
--1.1 have substantially lower [O/Fe] and higher [Na,Al/Fe] ratios, and 
2/3 of these stars also exhibit very high [La,Nd/Fe] ratios.  These abundance
variations will be discussed further in Section 5, but Figure \ref{f4} shows
that the light and heavy element abundance variations follow significant 
line strength variations and are not a result of model atmosphere 
uncertainties.

\section{RESULTS AND DISCUSSION}

Aside from studies by \citet[][high resolution; 1 star]{Pilachowski83} and
\citet[][medium resolution; 3 stars]{Wachter98}, which revealed that NGC 6229
exhibits an intrinsic [C/Fe] spread and is mildly $\alpha$--enhanced with
[Ca/Fe] = $+$0.2, little is known about the detailed chemical composition of
the cluster.  Therefore, the present work represents the first composition
analysis of a significant number (11 stars) of cluster members.

\subsection{Light Elements}

\begin{figure}
\epsscale{0.90}
\plotone{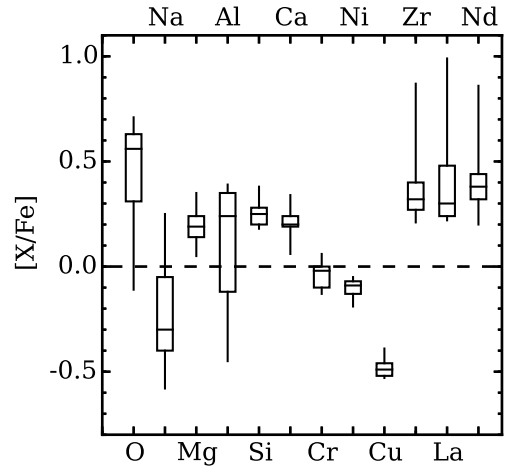}
\caption{A box plot summarizing the [X/Fe] distributions for all NGC 6229 stars
analyzed here.  For each element, the middle lines represent the median [X/Fe]
ratios, the box edges indicate the interquartile ranges, and the vertical
lines show the maximum and minimum abundances.}
\label{f5}
\end{figure}

\begin{figure*}
\epsscale{0.80}
\plotone{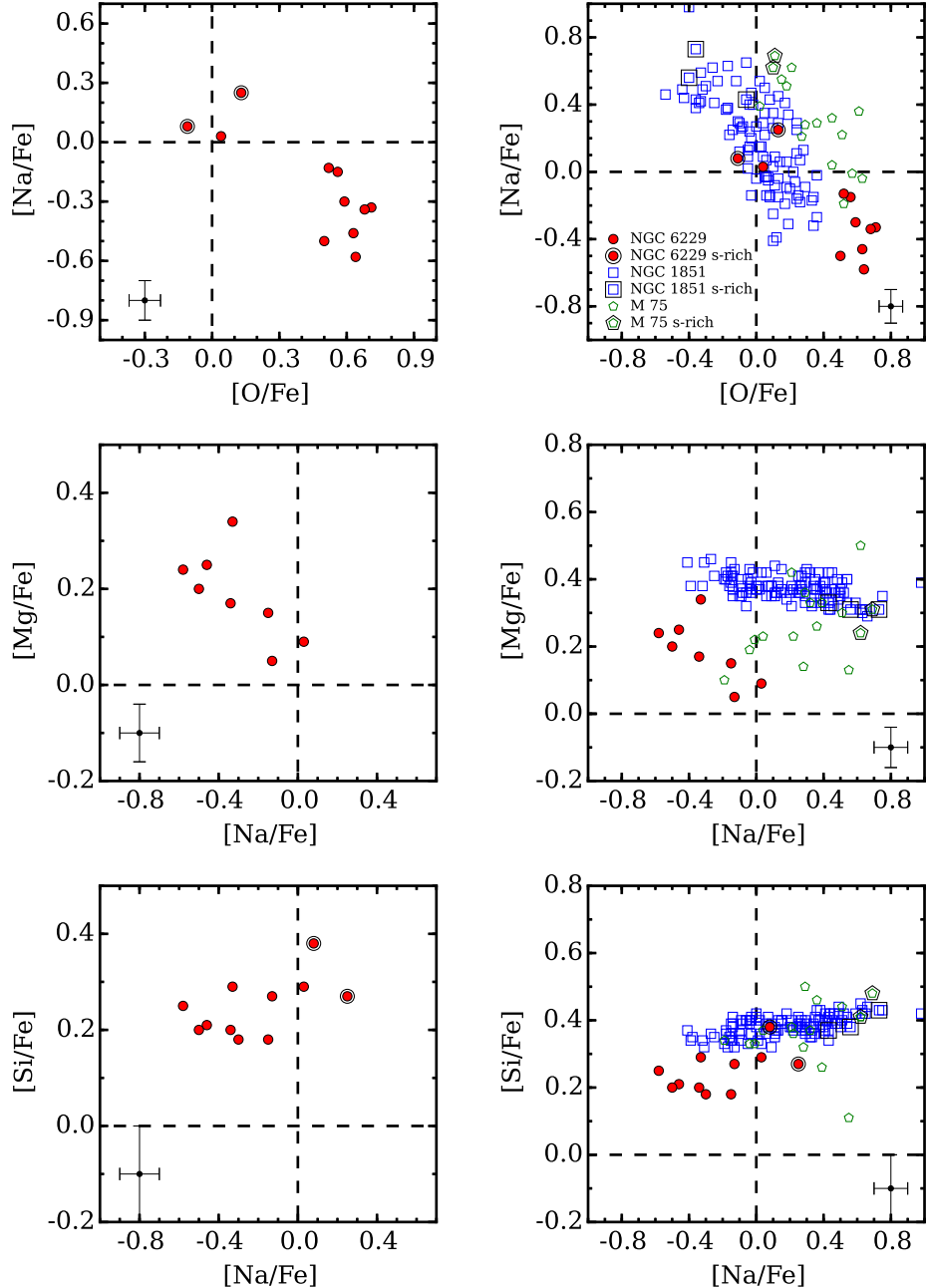}
\caption{Left: light element abundance variations between [O/Fe], [Na/Fe],
[Mg/Fe], and [Si/Fe] are shown for NGC 6229 (filled red circles).  The data
indicate that O--Na and Na--Mg are anti--correlated while [Na/Fe] and [Si/Fe]
exhibit a mild correlation.  A strong correlation between [Na/Fe] and [Al/Fe]
exists, but is omitted due to the small number of [Al/Fe] measurements.
Right: similar plots are shown comparing the light element variations between
NGC 6229, NGC 1851 \citep[open blue squares;][]{Carretta11}, and M 75
\citep[open green pentagons;][]{Kacharov13}.  For all three clusters, stars with
heavy element enhancements are indicated with large open symbols.}
\label{f6}
\end{figure*}

A summary of our derived [X/Fe] ratios is provided in Figure \ref{f5}, and 
shows in agreement with \citet{Wachter98} that the light elements exhibit large
star--to--star abundance variations.  In particular, [O/Fe], [Na/Fe], and
[Al/Fe] span full ranges of $\sim$0.85 dex, and Figure \ref{f6} indicates that
NGC 6229 exhibits the typical signatures of pollution from proton--capture
burning: O--Na and Na--Mg anti--correlations along with a mild Na--Si 
correlation.  Previous work has shown that the O--Na and Na--Al relations 
naturally arise when burning temperatures exceed $\sim$45--70 MK 
\citep[e.g.,][]{Langer97,Prantzos07}, but temperatures ranging from at least
70--100 MK are required in order to deplete \iso{24}{Mg} and produce 
\iso{28}{Si} \citep[e.g.,][]{Ventura11,Ventura13}.  Although the abundance 
variations of [Mg/Fe] and [Si/Fe] are significantly smaller in NGC 6229 than
those of [O/Fe] and the odd--Z elements, the clear Na--Mg and Na--Si 
(anti--)correlations presented in Figure \ref{f6} indicate that the cluster's
second generation stars must have formed from gas that was processed at
temperatures exceeding 70 MK.  However, we did not observe any correlations
between the light elements and [Ca/Fe], and NGC 6229 lacks the large [Ca/Mg]
spread that is characteristic of clusters that experienced more extensive
proton--capture nucleosynthesis \citep[e.g.,][]{Carretta15}.  We conclude that
the gas from which the second generation stars in NGC 6229 formed likely did 
not experience burning temperatures higher than $\sim$100 MK.

Although NGC 6229 possesses the classical light element (anti--)correlations,
its chemistry differs from most clusters in two key aspects: (1) both [Na/Fe]
and [Al/Fe] are abnormally low and (2) a large gap is present in the O--Na
anti--correlation (see Figure \ref{f6}).  The low [Na/Fe] and [Al/Fe] 
abundances are particularly interesting and will be discussed further in
Section 5.6, but their strong depletion likely indicates that NGC 6229 did not
form in its current environment.  The origin of the O--Na gap is not clear and
may simply reflect the small (11 star) sample size analyzed here; however, if
confirmed with a larger sample, NGC 6229 would possess one of the largest
O--Na discontinuities of any cluster.  Photometric studies have shown
that nearly all globular clusters contain stellar populations with unique
light element chemistry \citep[e.g.,][]{Piotto15,Milone17}, but the difference 
in [O/Fe] between adjacent populations is typically $\sim$0.3 dex or less 
\citep[e.g.,][]{Yong05,Marino08,Gratton12b,Carretta13,Kacharov13,Carretta15,
Johnson17b}.  In contrast, Figure \ref{f6} shows that the difference in [O/Fe]
between O--rich and O--poor stars for NGC 6229 may be as large as $\sim$0.5 dex.
Except for being shifted to lower [Na/Fe], the large O--Na gap in NGC 6229 
closely mirrors the distribution observed in $\omega$ Cen stars with --1.3 
$\la$ [Fe/H] $\la$ --0.9 \citep{Johnson10,Marino11a}, and suggests that the
two populations may have experienced similar formation processes\footnote{We 
note that both clusters also exhibit strikingly bimodal Na--Al correlations, 
but this distribution needs to be confirmed with a larger sample for 
NGC 6229.}.

The cluster's O--Na discontinuity may also reflect its ``clumpy" HB, which 
exhibits both a paucity of RR Lyrae stars and a gap in the blue HB 
\citep[e.g.,][]{Catelan98,Borissova99}.  Several studies have shown that a 
star's HB location is closely tied to its He abundance and light element 
composition \citep[e.g.,][]{Gratton11,Marino11c,Gratton12b,Villanova12,
Marino14,Gratton15}, and from Figure \ref{f6} one might expect that NGC 6229
is dominated by red HB stars.  However, the results from \citet{Borissova97} 
and \citet{Catelan98} indicate that the cluster has a ratio of red HB, RR 
Lyrae, and blue HB stars of about 32$\%$, 12$\%$, and 56$\%$, respectively.  
Therefore, NGC 6229 may be similar to clusters such as NGC 1851 and NGC 6723
where first generation stars, perhaps as a result of mild He enrichment or 
variable mass loss, populate both the red HB and a portion of the blue HB
\citep{Gratton12b,Gratton15}.  If the data presented here are representative
of the cluster's global first--to--second generation ratio, then the first
generation stars may also extend farther down the blue HB, similar to the
HB distribution predicted by \citet{Tailo16} for the O--rich/Na--poor 
intermediate metallicity stars of $\omega$ Cen.  In this case, the 27$\%$ 
(3/11) of stars in our sample that are clearly O--poor/Na--rich will evolve to 
become only the warmest HB and extreme HB stars.  A larger sample will be 
needed in order to determine if the O--Na gap is real and whether the number 
ratios of first and second generation stars can be reconciled with the 
cluster's HB morphology.

\begin{figure*}
\epsscale{1.00}
\plotone{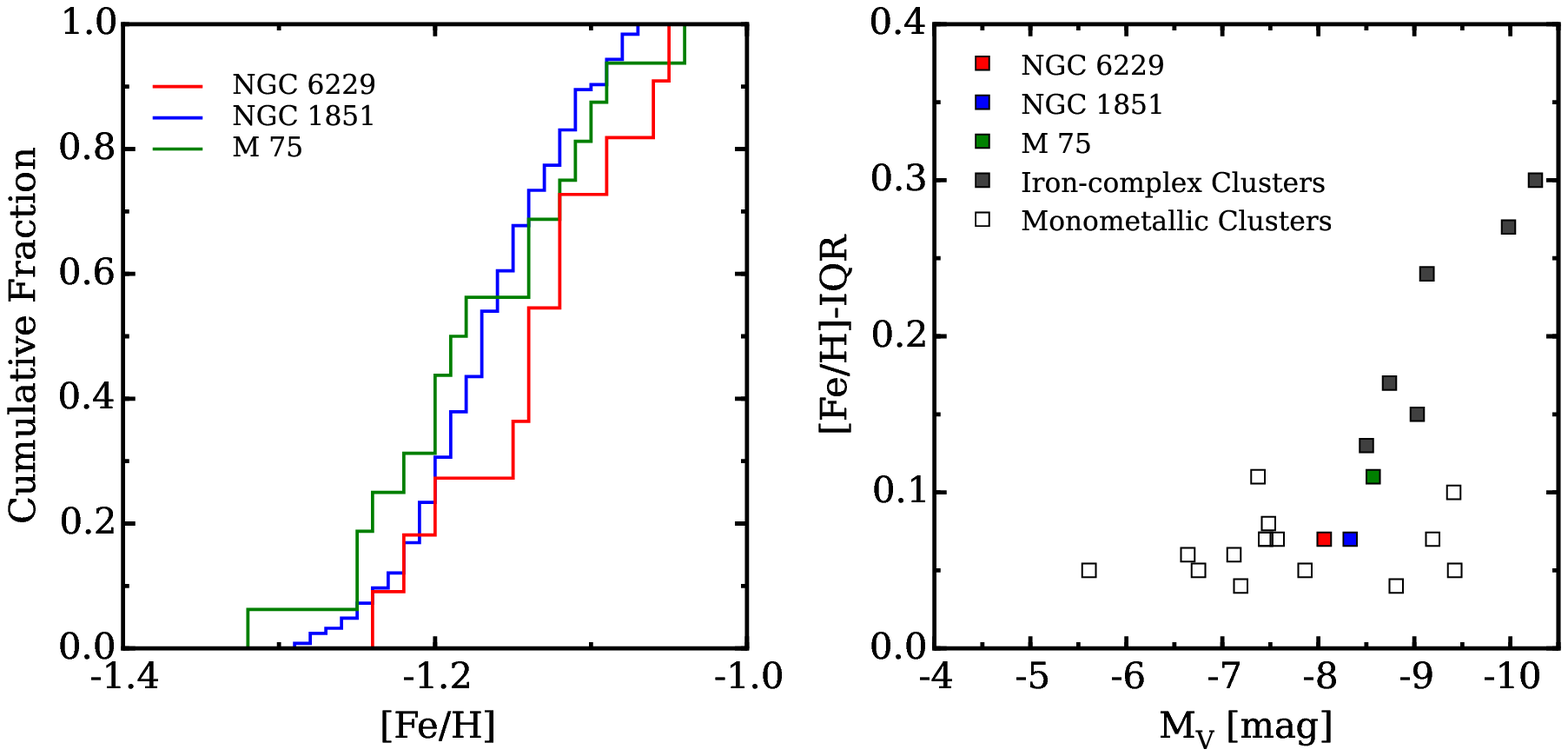}
\caption{Left: cumulative [Fe/H] distributions are shown for the
globular clusters NGC 6229 (this work), NGC 1851 \citep{Carretta11}, and
M 75 \citep{Kacharov13}.  Right: similar to Figure 1 in \citet{DaCosta16}, we
plot the measured [Fe/H] interquartile ranges of several monometallic (open
boxes) and iron--complex clusters (filled boxes) as a function of M$_{\rm V}$,
which we adopt as a proxy for mass.  The monometallic cluster data are from
\citet{Carretta09b}.  The iron--complex clusters, in addition to NGC 1851 and
M 75, include: $\omega$ Cen \citep{Johnson10}, M 54 \citep{Carretta10a},
NGC 6273 \citep{Johnson17a}, M 2 \citep{Yong14}, NGC 5286 \citep{Marino15},
and M 22 \citep{Marino11b}.  Note that the IQR for M 2 was set at 0.15 dex
because the [Fe/H] measurements provided by \citet{Yong14} trace a biased
sample.}
\label{f7}
\end{figure*}

\subsection{Heavy Elements}

For [Ca/Fe], [Cr/Fe], [Ni/Fe], and [Cu/Fe], the star--to--star dispersions
are $\sim$0.05 dex or less, and we find in agreement with \citet{Pilachowski83}
that the cluster is moderately $\alpha$--enhanced with 
$\langle$[Ca/Fe]$\rangle$ = $+$0.22 dex.  Additionally, [Cr/Fe] and [Ni/Fe] 
exhibit approximately Solar ratios, and [Cu/Fe] is strongly depleted with 
$\langle$[Cu/Fe]$\rangle$ = --0.48 dex.  In general, [Ca/Fe] is enhanced and
[Cu/Fe] is depleted in metal--poor globular clusters, but in Section 5.6 we 
provide some evidence that NGC 6229 may be slightly $\alpha$--poor and Cu--poor
compared to many similar metallicity Galactic clusters.

The heavy neutron--capture elements Zr, La, and Nd are all moderately enhanced
with mean [X/Fe] ratios of $\sim$0.4 dex and dispersions of $\sim$0.2 dex.
However, two stars in our sample have very high [La/Fe] and [Nd/Fe], and if
these stars are removed then the mean heavy element [X/Fe] ratios decrease to
$\sim$0.3 dex with dispersions of $\sim$0.1 dex.  These moderate heavy element
enhancements are typical for Galactic clusters near NGC 6229's metallicity 
\citep[e.g.,][]{James04}.  Interestingly, the two La/Nd--rich stars are also
among the most O--poor and Na/Al--rich in the cluster, which matches a pattern
observed in both M 75 \citep{Kacharov13} and NGC 1851 \citep[][see also Section
5.5]{Carretta11}.

Although we were unable to measure [Eu/Fe], we strongly suspect that the two
La/Nd--rich stars were polluted by material that experienced extensive
s--process enrichment, rather than the cluster exhibiting a large r--process
dispersion \citep[e.g.,][]{Roederer11}.  For example, we note above that 
NGC 6229 shares similar composition characteristics with M 75 and NGC 1851, 
and both clusters contain stars where the La/Nd--enhancements were driven by 
the s--process (i.e., the stars have high [La,Nd/Eu] ratios).  Furthermore, 
the La/Nd--enhanced stars in NGC 6229 would need to have [Eu/Fe] $\ga$ $+$1.2 
in order to maintain r--process dominated ratios of [La/Eu] $\sim$ --0.3, but 
even the most Eu--rich clusters do not host stars with [Eu/Fe] $\ga$ $+$1 
\citep[e.g.,][]{Johnson17b}.  Therefore, we conclude that NGC 6229 likely 
contains a separate population of stars that possess strong s--process 
enhancements.

\begin{figure*}
\epsscale{1.00}
\plotone{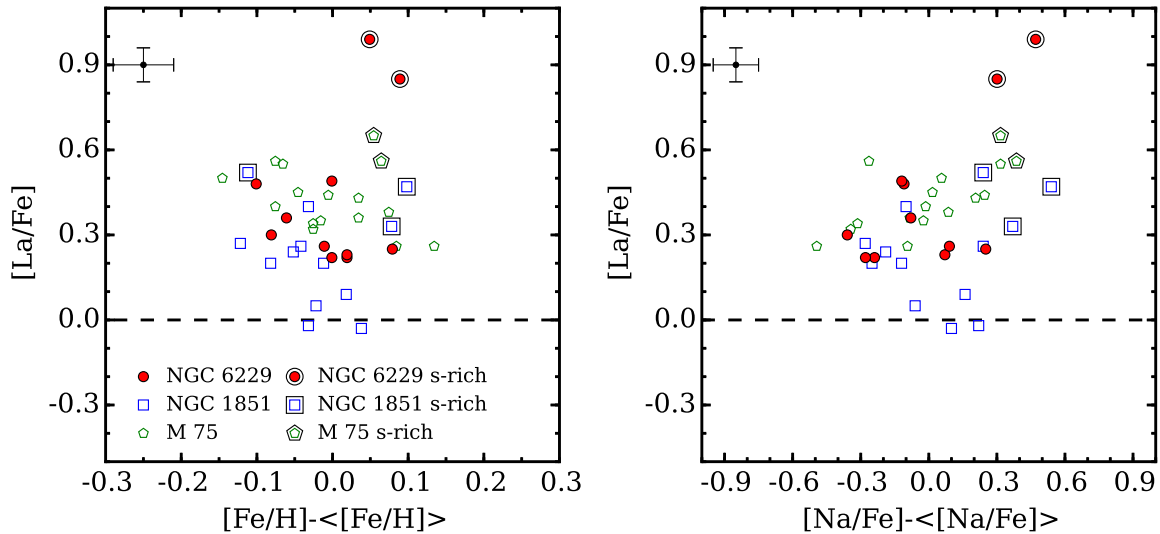}
\caption{Left: [La/Fe] abundances are plotted as a function of the difference
between a star's [Fe/H] value and the cluster mean for NGC 6229, NGC 1851, and
M 75.  The data sources are the same as in Figure \ref{f6}.  The star's with
larger symbols are those that were preferentially enriched by the s--process.
Right: a similar plot showing [La/Fe] as a function of the difference between
a star's [Na/Fe] abundance and the cluster mean.  The colors and symbols are
the same as in the left panel.  Note that the s--process enhanced stars tend
to exhibit both higher [Fe/H] and [Na/Fe] in all three clusters.  Error bars
representative of the values listed in Tables 4--5 are included in each
panel.}
\label{f8}
\end{figure*}

\subsection{Metallicity Distribution}

The few spectroscopic and photometric metallicity estimates that exist for
NGC 6229 span a wide range from [Fe/H] $\approx$ --1.5 to --1.05 dex, but 
the overall consensus is that the mean metallicity is probably around [Fe/H] =
--1.3 to --1.1 \citep{Pilachowski83,Carney91,Wachter98,Borissova99,Borissova01,
Arellano15}.  We note also that a reanalysis of the APOGEE data set by
\citet{Ness15} included 6 likely cluster members within 3$\arcmin$ of the 
cluster core and found $\langle$[Fe/H]$\rangle$ = --1.35 dex ($\sigma$ = 0.16 
dex).  However, the large dispersion is driven by one star (2MASS 
16464310+4731033) that is $\sim$0.35 dex more metal--rich than the main 
population.  For the present data set, we find $\langle$[Fe/H]$\rangle$ = 
--1.13 dex ($\sigma$ = 0.06 dex), which is higher than some studies but within 
the range estimated by previous investigators.  Similar to \citet{Ness15}, we 
also find 2MASS 16464310+4731033 to have a higher metallicity than most cluster
stars, but the difference is considerably smaller at $+$0.05 dex.  Despite the 
modest [Fe/H] dispersion measured here, the data suggest that NGC 6229 may 
possess light and heavy element signatures that are more similar to 
iron--complex, rather than monometallic, clusters.

\subsection{Monometallic or Iron--Complex?}

Since 18$\%$ (2/11) of the stars in our sample exhibit strongly enhanced
[La,Nd/Fe] ratios, it is prudent to investigate whether NGC 6229 is a lower
mass iron--complex cluster or a higher mass monometallic cluster.  Classifying
NGC 6229 as either a monometallic or iron--complex cluster primarily hinges on
three questions: (1) is the observed scatter in [Fe/H] significant?; (2) are
the enhancements in [La,Nd/Fe] correlated with [Fe/H]?; and (3) are the 
(presumably) s--process enhancements due to cluster pollution or binary mass 
transfer?  The first question is difficult to answer with a small sample, and
as stand--alone parameters the cluster's [Fe/H] dispersion (0.06 dex; 17$\%$) 
and interquartile range (IQR; 0.07 dex) values do not immediately flag the 
cluster as a member of the iron--complex class.  However, Figure \ref{f7} shows
that at M$_{\rm V}$ = --8.06 NGC 6229 spans a parameter space where the 
separation between iron--complex and monometallic clusters, based on 
metallicity spreads alone, is ambiguous.  

On the other hand, Figure \ref{f7} also indicates that except for small 
zero--point differences of $\sim$0.03 dex the metallicity distribution 
functions of NGC 6229, M 75, and NGC 1851 are nearly identical.  Critically, 
the latter two systems are recognized as iron--complex clusters 
\citep[][]{Marino15,DaCosta16} because their s--process enhanced stars are 
thought to form distinct \emph{populations}.  In both cases, the s--process
enhancements are correlated with small ($\sim$0.05--0.10 dex) increases in 
[Fe/H] \citep{Carretta11,Kacharov13}.  Figure \ref{f8} shows that NGC 6229
exhibits the same trend with the two La/Nd--rich stars being $\sim$0.05--0.10
dex more metal--rich than the cluster average.  When all three clusters are 
combined, Figure \ref{f8} shows that we obtain a stronger correlation
signature between [Fe/H] and [La,Nd/Fe] enhancements.  In fact, a set of 
10$^{\rm 6}$ random samplings of the combined data indicates only a 0.45$\%$ 
chance that one would draw 6 or more La/Nd--rich stars with [Fe/H] values at 
least 0.05 dex higher than the cluster average, and we conclude for all three 
clusters that the heavy element enhancements are likely correlated with small,
but real, [Fe/H] enhancements.

An alternative explanation for the La/Nd--rich stars in NGC 6229 and other 
clusters, especially those for which only small samples of enriched stars have
been found, is that they experienced mass transfer from binary companions.
For NGC 6229, we did not detect radial velocity variations over the 3 month
observing window, but cannot directly test the mass transfer scenario further
without measuring additional heavy elements \citep[e.g., see][]{Roederer16}.
However, previous work indicates that mass transfer alone is unlikely to 
produce significant numbers of s--process enhanced stars in globular clusters, 
and we note that \citet{D'Orazi10} estimate that the binary fraction among
second generation (Na--rich) cluster stars is only $\sim$1$\%$.  In this 
context, we note that Figures \ref{f6} and \ref{f8} show that both La/Nd--rich 
stars in our sample are very Na--rich, and it is unlikely that a random 
sampling of 11 RGB stars would yield such a comparatively high fraction 
of second generation stars enriched by mass transfer events.  Instead, a 
cluster self--enrichment process is the more probable scenario.  Therefore, we 
conclude that NGC 6229 likely contains a separate population of stars enriched 
by the s--process and is the lowest mass iron--complex cluster currently known 
in the Galaxy\footnote{We note that \citet{Simmerer13} found a potential [Fe/H]
spread in NGC 3201, and that the cluster has a fainter absolute magnitude than 
NGC 6229 with M$_{\rm V}$ = --7.45 \citep{Harris96}.  However, 
\citet{Mucciarelli15} disputes the [Fe/H] spread claim, and we note that 
\citet{Munoz13} did not find any stars with strong s--process enhancements.  
Therefore, even if NGC 3201 does have a metallicity spread it may not share 
the same chemical traits that link other iron--complex clusters.}.

\begin{figure*}
\epsscale{0.35}
\plotone{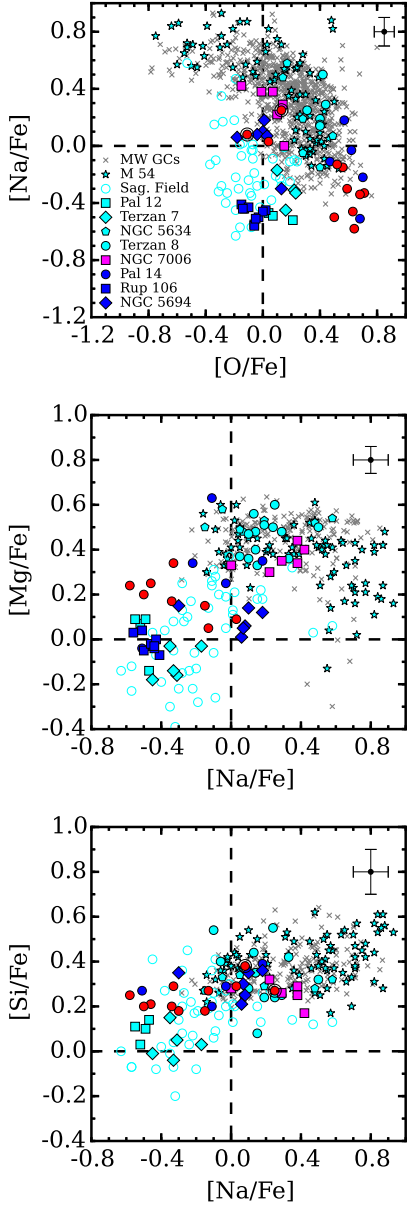}
\caption{Similar to Figure \ref{f6}, the light element abundance trends of
NGC 6229 (filled red circles) are compared against those of several Milky Way
clusters.  The light grey crosses indicate data from \citet{Carretta09b,
Carretta09c}, and represent the general trends exhibited by $\sim$15
monometallic clusters.  The cyan symbols represent the abundance trends of
several clusters (filled symbols) and field stars (open circles) associated
with the Sagittarius dwarf spheroidal galaxy \citep{Bonifacio00,Cohen04,
Sbordone07,Carretta10a,McWilliam13,Carretta14b,Carretta17}.  The filled magenta
boxes indicate the abundance trends of NGC 7006 \citep{Kraft98}, which has a
tentative association with NGC 6229.  The filled blue symbols represent
extragalactic halo clusters that have been accreted by the Milky Way
\citep{Caliskan12,Mucciarelli13,Villanova13}.}
\label{f9}
\end{figure*}

\subsection{Implications of Correlated Light and Heavy Element Abundance 
Variations}

In addition to sharing similar metallicities and heavy element patterns, 
NGC 6229, M 75 \citep{Kacharov13}, and NGC 1851 \citep{Yong08,Yong09,
Carretta11,Gratton12c,Carretta14a,Simpson17} also exhibit an interesting 
correlation between their light and heavy element abundances that may provide
insight into their formation.  Figure \ref{f8} indicates that the s--process
enhanced stars are consistently among the most Na--rich in their 
respective clusters despite the fact that all three systems exhibit different
light element (anti--)correlations\footnote{\citet{Carretta11} showed that the
metal--poor and metal--rich populations in NGC 1851 contain stars with similar 
light element abundance ranges.  However, in the context of this paper the 
correlation between light and heavy elements in NGC 1851 refers to the strong 
preference of stars with [Ba/H] $>$ --0.4 to have enhanced [Na/Fe].  For 
example, NGC 1851 stars with [Ba/H] $\leq$ --0.4 have $\langle$[Na/Fe]$\rangle$
= $+$0.14 dex ($\sigma$ = 0.26 dex) while those with higher [Ba/H] have 
$\langle$[Na/Fe]$\rangle$ = $+$0.51 dex ($\sigma$ = 0.26 dex).  In other words,
the s--process enhanced population in NGC 1851 is dominated by second 
generation (Na--rich) stars.}.  For example, Figure \ref{f6} shows that NGC 
6229 has significantly lower [Na/Fe] and [Si/Fe] than M 75 and NGC 1851, M 75 
exhibits a smaller range in [O/Fe] than NGC 6229 and NGC 1851, and only 
NGC 6229 and NGC 1851 have clear Na--Mg anti--correlations.  Interestingly, we 
note that, in addition to sharing a large gap in the O--Na relation with NGC 
6229, the intermediate metallicity (--1.3 $\la$ [Fe/H] $\la$ --0.9) stars of 
$\omega$ Cen also exhibit a similar preference for the s--process enhanced 
stars to have high [Na/Fe] \citep{Johnson10,Marino11a}.  Furthermore, the 
$\omega$ Cen stars likely experienced even more extreme proton--capture 
processing than NGC 6229, M 75, or NGC 1851.  However, similar light and heavy 
element correlations are not present in the more metal--poor $\omega$ Cen stars
nor in any other iron--complex clusters \citep{Marino09,Yong14,Marino15,
Roederer16,Johnson17a}, and s--process production is not correlated with light 
element abundance variations in most monometallic clusters 
\citep[e.g.,][]{D'Orazi10}.  

\begin{figure*}
\epsscale{0.60}
\plotone{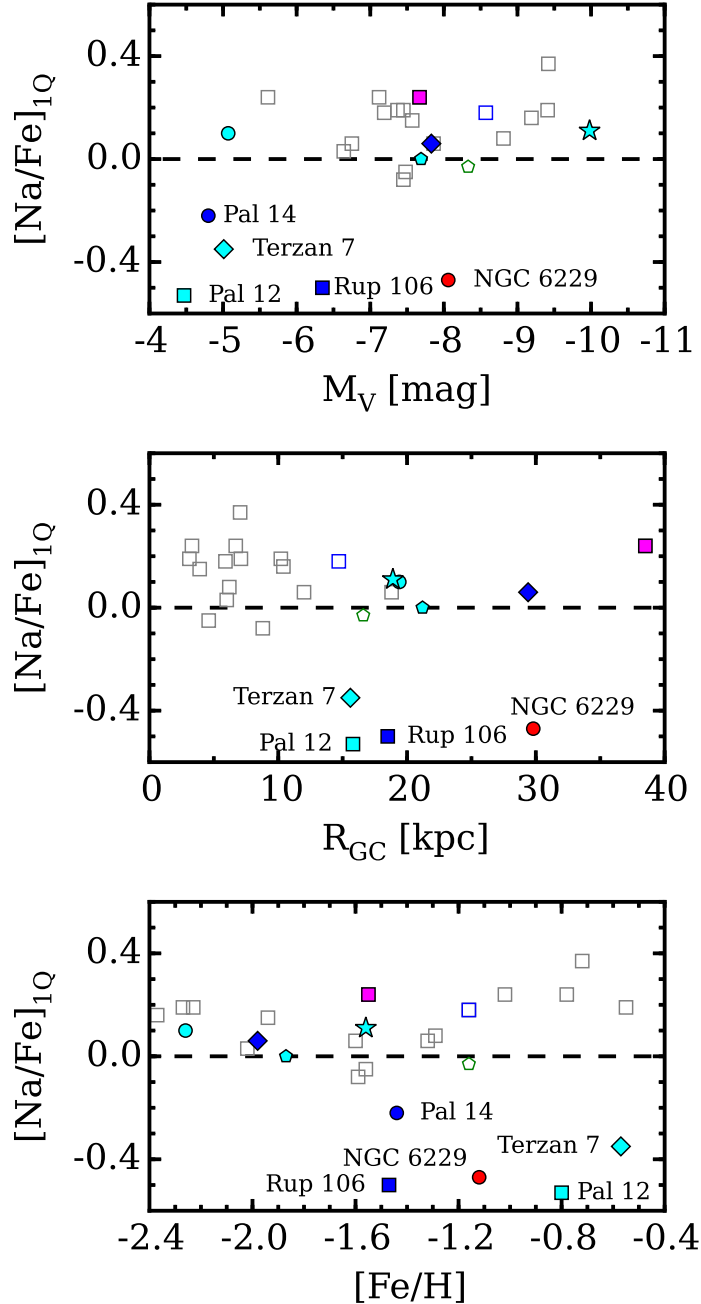}
\caption{The top, middle, and bottom panels plot the first quartile [Na/Fe]
abundances ([Na/Fe]$_{\rm 1Q}$) of several globular clusters as functions of
M$_{\rm V}$, Galactocentric distance (R$_{\rm GC}$), and mean [Fe/H],
respectively.  The M$_{\rm V}$ and R$_{\rm GC}$ data are from \citet{Harris96},
and the literature abundances, colors, and symbols are the same as those in
Figures \ref{f6}--\ref{f9}.  Note that all clusters with [Na/Fe]$_{\rm 1Q}$
$\la$ --0.1, which are identified by name, are either associated with the
Sagittarius dwarf spheroidal system (cyan symbols) or are strongly suspected
to have been accreted by the Milky Way (blue symbols).}
\label{f10}
\end{figure*}

The data suggest that NGC 6229, M 75, NGC 1851, and the intermediate 
metallicity $\omega$ Cen stars all experienced similar modes of star formation,
even if the chemical enrichment details (e.g., polluter mass ranges; initial 
compositions) varied from cluster--to--cluster.  In this light, the delayed 
binary Type II supernova model outlined by \citet{D'Antona16} and 
\citet{D'Ercole16} provides a context with which we may interpret formation 
differences between clusters such as NGC 6229, which have Fe/s--process 
enhanced populations dominated by second generation stars, and clusters such as
M 22, which have Fe/s--process enhanced populations exhibiting a mixture of 
first and second generation stars.  

\begin{figure*}
\epsscale{0.80}
\plotone{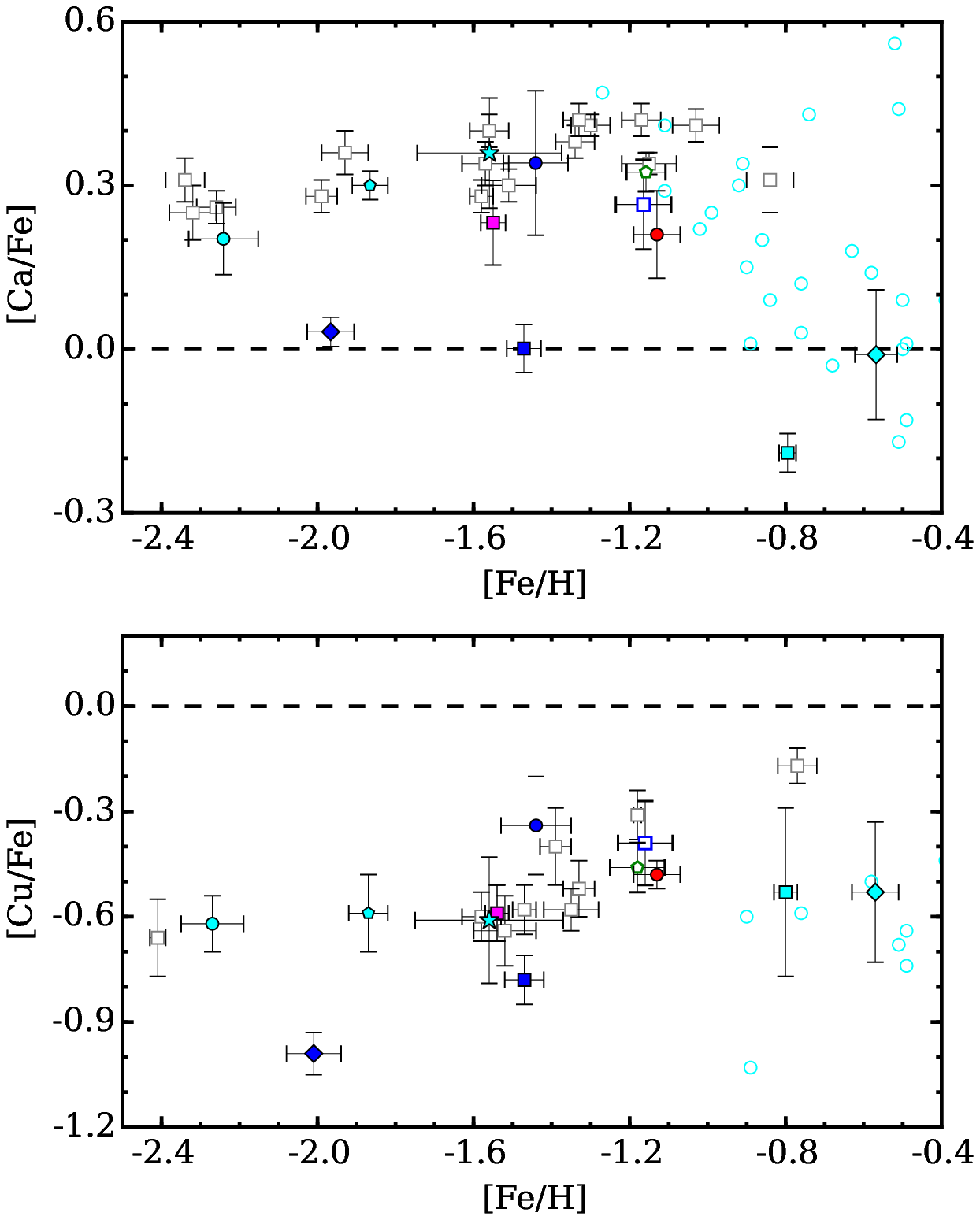}
\caption{Top: [Ca/Fe] abundances of several Galactic globular clusters are
plotted as a function of [Fe/H].  Except for the Sagittarius dwarf field stars,
for which the open cyan circles represent individual stars, each symbol
indicates the mean [Ca/Fe] and [Fe/H] values for a cluster.  Similarly, the
error bars indicate the [Ca/Fe] and [Fe/H] dispersions of each cluster.  The
open grey boxes represent clusters from \citet{Carretta10b}, and the
remaining colors, symbols, and references are the same as in
Figures \ref{f6}--\ref{f9}.  Bottom: Similar to the top panel, mean [Cu/Fe]
abundances of several Milky Way clusters are plotted as a function of [Fe/H].
The reference Galactic cluster population (open grey boxes) data are from
\citet{Simmerer03}.  For both panels, the color scheme follows the outline
of Figure \ref{f9} where cyan symbols indicate objects associated with the
Sagittarius dwarf spheroidal galaxy, filled magenta symbols represent objects
with a claimed association to NGC 6229, and filled blue symbols indicate
clusters accreted by the Milky Way.}
\label{f11}
\end{figure*}

For both cases listed above, \citet{D'Antona16} and \citet{D'Ercole16} suggest
that a mixing of AGB ejecta with pristine gas produces the light element 
abundance variations present in the ``metal--poor" populations of 
iron--complex clusters.  However, if a sufficient number of massive stars are 
initially below the supernova threshold ($\sim$8--10 M$_{\rm \odot}$), are in
close binary systems, and experience enough mass transfer to undergo core 
collapse, then a second ``delayed" supernova epoch may occur and form a new
population of stars with higher [Fe/H].  In this scenario, 
\citet{D'Ercole16} notes that inhomogeneous mixing between supernova and AGB
gas could produce both first and second generation stars when the mixed gas 
is reaccreted by the cluster, which would result in a system resembling M 22.
Therefore, we speculate that the more metal--rich populations present in 
clusters similar to NGC 6229, where the ratio of second to first generation
stars is considerably higher than in the metal--poor populations, could have 
formed if the delayed supernova gas and AGB ejecta were instead well--mixed.
Alternatively, metal--rich first and second generations stars could form at 
different times, but this would require second generation stars to form first, 
perhaps as a result of a radial composition gradient, and also for star 
formation to end at earlier times in lower mass iron--complex clusters.  
Differences in radiative cooling time could also help explain a scenario in 
which second generation stars form more efficiently in higher metallicity 
populations \citep[e.g., ][see their Section 4.2.3]{Herwig12}.  We note that 
cluster mergers may also remain a viable explanation 
\citep[e.g., ][]{Carretta10c,Gavagnin16}.

\subsection{Clues to NGC 6229's Origin and Evolution}

NGC 6229 shares a variety of chemical characteristics with both monometallic 
and iron--complex clusters, but the cluster's low [Na/Fe] (and [Al/Fe])
abundances distinguish it from most other Galactic systems.  Figures 
\ref{f9}--\ref{f10} indicate that NGC 6229 shares its low [Na/Fe] ratios with 
the clusters Rup 106, Pal 12, Pal 14, and Terzan 7, and all four systems are 
strongly suspected to have extragalactic origins.  For example, Pal 12 
\citep{Cohen04} and Terzan 7 \citep{Sbordone07} are strongly tied to the 
Sagittarius stream while Pal 14 \citep{Caliskan12} and Rup 106 
\citep{Villanova13} exhibit chemical signatures reminiscent of those found in
dwarf galaxies and/or their globular cluster populations.  Figure \ref{f11} 
also indicates that NGC 6229 may be deficient by $\sim$0.2 dex in [Ca/Fe] and 
[Cu/Fe] compared to similar metallicity halo clusters.  We note that the 
combination of low [Na/Fe], [$\alpha$/Fe], and [Cu/Fe], particularly at higher 
metallicity, is a chemical signature that is frequently associated with 
accreted systems, including $\omega$ Cen \citep{Cunha02,Pancino02}, field 
stars within the Sagittarius dwarf spheroidal galaxy 
\citep[e.g.,][]{McWilliam13}, and accreted Galactic field stars 
\citep[e.g.,][]{Nissen11}.  Combining these data with NGC 6229's $\sim$30 kpc 
galactocentric radius leads us to conclude that NGC 6229 is also an accreted 
cluster and probably formed in a more massive progenitor system that has since 
been tidally disrupted by the Milky Way.

Previous attempts to match NGC 6229 to known tidal streams have been 
unsuccessful \citep[e.g.,][]{Grillmair14}, but the cluster may have been part
of a relatively low mass system and/or was accreted early in the Galaxy's 
history.  Unfortunately, little is known about the cluster's orbit, but we 
can note that NGC 6229 may share a similar perigalactic distance ($\sim$4 kpc)
with NGC 5694 \citep{vandenBergh95}, which has been shown to have an 
extragalactic origin \citep{Mucciarelli13}.  \citet{Palma02} suggested that 
NGC 6229 may share a connection with NGC 7006, Pal 4, and Pyxis; however, the 
data presented here (see Figures \ref{f9}--\ref{f10}) and in \citet{Kraft98}
do not provide strong evidence supporting a chemical link between NGC 6229 and 
NGC 7006.  Additional observations of individual stars, along with measurements
of cluster orbital parameters, are required to determine if NGC 6229, Pal 4,
and/or Pyxis share a common origin.

Even though NGC 6229 exhibits some chemical properties associated with accreted
clusters, Figures \ref{f9}--\ref{f10} also highlight significant differences.
For example, all of the low--Na ([Na/Fe]$_{\rm 1Q}$\footnote{[Na/Fe]$_{\rm 1Q}$
represents the first quartile of a cluster's [Na/Fe] distribution.} $<$ --0.1)
clusters have [Fe/H] $\ga$ --1.5, but NGC 6229 is both significantly more
massive and resides at a larger galactocentric radius than the other accreted
systems (except Pal 14).  Figure \ref{f9} also indicates that NGC 6229 has a
more extended O--Na anti--correlation than other accreted clusters, and is the
only low--Na cluster to exhibit both a clear Na--Mg anti--correlation and 
Na--Si correlation.  The more extensive light element variations in NGC 6229
are consistent with its high mass and extended blue HB, and suggest that the
cluster experienced pollution from higher temperature (mass?) sources than its 
lower mass low--Na counterparts.

In this light, the heavy element abundance spread and extensive light element 
variations of NGC 6229 more closely resemble the composition characteristics
of M 54 \citep[e.g.,][]{Carretta10a}, the iron--complex cluster and likely
nuclear star cluster of the Sagittarius dwarf spheroidal galaxy, than the lower
mass Sagittarius clusters Pal 12 and Terzan 7.  As noted by \citet{DaCosta16},
the similar chemical properties shared by M 54 and other iron--complex clusters
is a key component of the hypothesis that these systems are former dwarf galaxy
nuclei.  By extension, we posit that NGC 6229 was not only accreted by the
Milky Way, but may have been the nuclear star cluster of a system similar to,
but less massive than, the Sagittarius dwarf spheroidal galaxy.  However, the
low [Na,Al/Fe] abundances of NGC 6229 indicate that the cluster, and by
extension any progenitor system, experienced a different enrichment history
than Sagittarius.

\section{SUMMARY}

Using high resolution (R $\approx$ 38,000) and S/N ($>$ 100) spectra obtained
with the MMT--Hectochelle multi--fiber spectrograph, we have identified 14
(out of a total sample of 57 targets) bright RGB member stars located near the 
outer halo globular cluster NGC 6229.  We measured a mean cluster heliocentric 
radial velocity of --138.1$_{-1.0}^{+1.0}$ km s$^{\rm -1}$ and a modest velocity
dispersion of 3.8$_{-0.7}^{+1.0}$ km s$^{\rm -1}$.  The data suggest that
NGC 6229 has a low, but not extreme, mass--to--light ratio of 
(M/L$_{\rm V}$)$_{\rm \odot}$ = 0.82$_{-0.28}^{+0.49}$.

For 11/14 member stars, we were able to perform a detailed chemical composition
analysis using several elements ranging from oxygen to neodymium.  We found
that NGC 6229 exhibits the classical light element abundance variations 
associated with globular clusters, including clear O--Na and Na--Mg 
anti--correlations and Na--Al and Na--Si correlations.  The light element 
abundance data indicate that the pollution sources within NGC 6229 must have 
reached temperatures of at least 75 MK, but the paucity of additional 
correlations involving elements heavier than Si likely sets an upper limit to 
the burning conditions of $\sim$100 MK.  The light elements are 
particularly interesting for NGC 6229 because the first and second generation
stars are separated by a gap of $\sim$0.5 dex in [O/Fe] and the [Na,Al/Fe]
ratios are unusually low.

In general, NGC 6229 appears to be moderately $\alpha$--enhanced with a mean
[Ca/Fe] = $+$0.22 dex ($\sigma$ = 0.08 dex).  Additionally, the Fe--peak 
elements exhibit approximately Solar [X/Fe] ratios with dispersions of 
$\sim$0.05 dex.  Similar to other metal--poor clusters, we found copper to be 
strongly depleted with $\langle$[Cu/Fe]$\rangle$ = --0.48 dex ($\sigma$ = 0.04 
dex); however, NGC 6229 may be deficient by $\sim$0.2 dex in both [Cu/Fe] and 
[Ca/Fe] compared to other similar metallicity halo clusters.  The 
neutron--capture elements Zr, La, and Nd were found to be uniformly enhanced 
by $\sim$0.3 dex with small star--to--star dispersions, but 2/11 stars 
exhibit strongly enhanced [La/Fe] and [Nd/Fe] ratios that are correlated
with a small ($\sim$0.05 dex) increase in [Fe/H].  Although we were not able to 
measure an r--process indicator such as [Eu/Fe], we strongly suspect that the 
[La,Nd/Fe] enhancements were due to s--process enrichment.

Previous investigations measured a variety of metallicities ranging from [Fe/H]
$\approx$ --1.50 to --1.05 dex, and we found NGC 6229 to be near the higher 
end of this range with $\langle$[Fe/H]$\rangle$ = --1.13 dex ($\sigma$ = 0.06 
dex).  Interestingly, NGC 6229 was found to share a variety of peculiar 
chemical signatures with the similar metallicity iron--complex clusters 
M 75 and NGC 1851, and also with the intermediate metallicity populations of
$\omega$ Cen.  In particular, all four systems exhibit: (1) distinct
populations of stars with s--process enhancements; (2) similar 
correlations between [Fe/H] and [La,Nd/Fe]; and (3) unusual correlations
between [Na/Fe] and [La,Nd/Fe].  Although the [Fe/H] spread alone
is insufficient to identify NGC 6229 as an iron--complex cluster, the 
chemical signatures listed above combined with the fact that M 75, NGC 1851, 
and $\omega$ Cen have been identified as iron--complex systems lead us to 
conclude that NGC 6229 is also an iron--complex cluster.  If confirmed with a
larger sample, NGC 6229 would be the lowest mass iron--complex cluster known
in the Galaxy.

Of the chemical signatures listed above, we found the consistent correlation 
between [Na/Fe] and [La,Nd/Fe] among several clusters to be particularly 
interesting.  We interpreted this trend within the context of the delayed 
Type II supernova model discussed in \citet{D'Antona16} and \citet{D'Ercole16},
and speculated that the predominance of Na--rich stars in the Fe/s--process
enhanced populations of clusters such as NGC 6229 may be driven by homogeneous 
mixing between delayed supernova gas and AGB ejecta.  However, differences in 
star formation efficiency and time, perhaps aided by variations in the cooling
properties of first and second generation star gas, and also cluster merger 
scenarios, remain possible explanations of why the Fe/s--process enhanced
populations of some clusters are dominated by second generation stars and
others have a mixture of first and second generation stars.

The signature of low [Na/Fe], [Al/Fe], [Ca/Fe], and [Cu/Fe]
in NGC 6229 strongly suggests that the cluster has an extragalactic origin,
but no clear progenitor galaxy has been identified.  Previous work noted a
possible association between NGC 6229 and NGC 7006, but the data presented here
and in the literature do not support a strong chemical link between the 
clusters.  NGC 6229 has also been linked to Pal 4 and Pyxis, but additional
composition analyses of individual stars are required before any clear 
associations can be made.  Although several accreted clusters in the Milky Way 
have low [Na/Fe], NGC 6229 seems to stand out as having experienced more 
extensive chemical enrichment and is considerably more massive.  In fact, 
the chemical and physical characteristics of NGC 6229 may be a closer match to 
M 54 than the lower mass clusters accreted from the Sagittarius dwarf 
spheroidal galaxy.  Therefore, we speculate that NGC 6229 may also be the 
former core of a system that was similar to, but less massive than, the 
Sagittarius galaxy.  However, the cluster's low [Na,Al/Fe] abundances 
indicate that NGC 6229 and any progenitor system would have experienced a 
different chemical enrichment history than Sagittarius.

Future work may focus on obtaining a deeper understanding of the connection
between clusters such as NGC 6229, M 75, and NGC 1851, which share several
chemical characteristics but may have very different sub--giant branch (SGB)
morphologies.  In particular, it would be instructive to know whether NGC 6229
has a clear double SGB similar to NGC 1851.  Additionally, it will be 
especially important to obtain larger spectroscopic samples of stars in NGC 
6229, M 75, and other outer halo clusters to confirm whether these systems 
truly host significant populations of s--process enhanced stars.  For NGC 6229,
[Eu/Fe] measurements will help confirm that the [La,Nd/Fe] enhancements 
observed here are driven by the s--process, and a larger sample of [O/Fe] and 
[Na/Fe] abundances will show whether or not the large gap in the O--Na relation
is real.  An updated orbital solution for NGC 6229 would also be helpful
for investigating further any connections between the cluster and stellar 
streams, and will also help trace its origin.  

\acknowledgements

This research has made use of NASA's Astrophysics Data System Bibliographic
Services.  This publication has made use of data products from the Two Micron
All Sky Survey, which is a joint project of the University of Massachusetts
and the Infrared Processing and Analysis Center/California Institute of
Technology, funded by the National Aeronautics and Space Administration and
the National Science Foundation.  This paper uses data products produced by the
OIR Telescope Data Center, supported by the Smithsonian Astrophysical 
Observatory.  C.I.J. gratefully acknowledges support from the Clay Fellowship, 
administered by the Smithsonian Astrophysical Observatory.  R.M.R acknowledges 
support from grant AST--1413755 from the National Science Foundation.  M.G.W. 
is supported by National Science Foundation grants AST--1313045 and 
AST--1412999.

\clearpage
\tablenum{1}
\tablecolumns{8}
\tablewidth{0pt}

\begin{deluxetable}{cccccccc}
\tabletypesize{\tiny}
\tablecaption{Star Identifiers, Coordinates, Photometry, and Velocities}
\tablehead{
\colhead{Star Name}     &
\colhead{RA}      &
\colhead{DEC}      &
\colhead{J}      &
\colhead{H}      &
\colhead{K$_{\rm S}$}      &
\colhead{RV$_{\rm helio.}$}      &
\colhead{RV$_{\rm helio.}$ Error}       \\
\colhead{(2MASS)}	&
\colhead{(degrees)}      &
\colhead{(degrees)}      &
\colhead{(mag.)}      &
\colhead{(mag.)}      &
\colhead{(mag.)}      &
\colhead{(km s$^{\rm -1}$)}      &
\colhead{(km s$^{\rm -1}$)}      
}

\startdata
\hline
\multicolumn{8}{c}{Cluster Members}       \\
\hline
16464310+4731033	&	251.679603	&	47.517605	&	13.870	&	13.142	&	13.027	&	$-$139.6	&	0.4	\\
16465095+4731163	&	251.712295	&	47.521198	&	12.482	&	11.659	&	11.492	&	$-$136.4	&	0.6	\\
16465319+4732413	&	251.721631	&	47.544823	&	12.462	&	11.650	&	11.520	&	$-$140.1	&	0.5	\\
16465362+4730222	&	251.723456	&	47.506184	&	13.383	&	12.649	&	12.493	&	$-$135.1	&	0.3	\\
16465422+4733371	&	251.725949	&	47.560307	&	12.315	&	11.463	&	11.283	&	$-$145.6	&	0.6	\\
16465515+4731485	&	251.729808	&	47.530163	&	12.761	&	12.022	&	11.867	&	$-$134.8	&	0.6	\\
16465763+4732296	&	251.740135	&	47.541565	&	13.763	&	13.147	&	13.015	&	$-$138.9	&	0.2	\\
16470094+4731065	&	251.753932	&	47.518482	&	14.315	&	13.592	&	13.540	&	$-$132.2	&	0.5	\\
16470210+4733163	&	251.758753	&	47.554543	&	13.869	&	13.153	&	13.106	&	$-$138.5	&	0.4	\\
16471104+4728244	&	251.796019	&	47.473461	&	13.203	&	12.499	&	12.312	&	$-$140.4	&	0.9	\\
16471308+4732430	&	251.804508	&	47.545284	&	13.749	&	13.114	&	12.998	&	$-$144.0	&	0.3	\\
16472079+4731478	&	251.836647	&	47.529964	&	14.158	&	13.497	&	13.332	&	$-$136.0	&	0.4	\\
16470403+4731201	&	251.766795	&	47.522274	&	14.852	&	14.193	&	14.158	&	$-$140.0	&	0.4	\\
16470289+4732471	&	251.762075	&	47.546425	&	14.099	&	13.485	&	13.446	&	$-$136.0	&	0.3	\\
\hline
\multicolumn{8}{c}{Non$-$Members}       \\
\hline
16460104+4747575	&	251.504343	&	47.799309	&	14.189	&	13.566	&	13.533	&	$-$15.0	&	0.6	\\
16463345+4741560	&	251.639382	&	47.698902	&	14.055	&	13.428	&	13.287	&	$-$25.1	&	0.4	\\
16461569+4743371	&	251.565412	&	47.726974	&	14.530	&	13.901	&	13.758	&	$-$26.3	&	0.4	\\
16470381+4746477	&	251.765914	&	47.779942	&	13.512	&	12.878	&	12.728	&	$-$90.2	&	0.6	\\
16463660+4745098	&	251.652502	&	47.752739	&	14.763	&	14.113	&	14.015	&	$-$54.3	&	0.6	\\
16461166+4745117	&	251.548617	&	47.753258	&	14.585	&	13.914	&	13.815	&	$-$14.1	&	0.9	\\
16461965+4744207	&	251.581876	&	47.739101	&	14.125	&	13.514	&	13.458	&	$-$36.5	&	0.7	\\
16465487+4727249	&	251.728632	&	47.456921	&	12.325	&	11.669	&	11.480	&	$-$5.1	&	0.2	\\
16465059+4725208	&	251.710830	&	47.422451	&	14.486	&	13.927	&	13.790	&	$-$96.8	&	0.5	\\
16472581+4732161	&	251.857562	&	47.537819	&	13.905	&	13.558	&	13.560	&	$-$6.4	&	0.3	\\
16471626+4733136	&	251.817755	&	47.553802	&	14.502	&	13.921	&	13.911	&	$-$3.7	&	0.3	\\
16464036+4735195	&	251.668207	&	47.588772	&	14.602	&	13.932	&	13.800	&	$-$30.0	&	0.5	\\
16464167+4730049	&	251.673662	&	47.501385	&	13.585	&	13.305	&	13.146	&	7.9	&	0.3	\\
16465290+4728328	&	251.720420	&	47.475780	&	13.767	&	13.215	&	13.061	&	$-$30.3	&	0.5	\\
16465601+4730407	&	251.733386	&	47.511307	&	14.452	&	13.752	&	13.663	&	$-$11.3	&	0.5	\\
16473405+4744203	&	251.891907	&	47.738995	&	13.494	&	12.895	&	12.699	&	$-$25.0	&	0.3	\\
16472302+4737244	&	251.845944	&	47.623466	&	14.331	&	13.843	&	13.704	&	$-$5.1	&	0.2	\\
16470431+4734268	&	251.767967	&	47.574135	&	14.378	&	13.984	&	13.890	&	$-$58.5	&	0.4	\\
16470931+4736507	&	251.788805	&	47.614105	&	14.420	&	13.782	&	13.665	&	$-$43.2	&	0.4	\\
16471359+4734059	&	251.806635	&	47.568306	&	14.568	&	14.029	&	13.816	&	$-$0.4	&	0.6	\\
16471388+4735472	&	251.807857	&	47.596451	&	13.086	&	12.618	&	12.542	&	$-$38.1	&	0.2	\\
16470843+4744032	&	251.785164	&	47.734230	&	14.818	&	14.269	&	14.140	&	$-$78.3	&	33.3	\\
16484573+4736346	&	252.190551	&	47.609638	&	14.787	&	14.169	&	14.062	&	$-$6.4	&	0.7	\\
16482177+4717330	&	252.090731	&	47.292526	&	14.653	&	14.005	&	13.890	&	$-$99.4	&	0.3	\\
16482482+4718547	&	252.103440	&	47.315212	&	14.309	&	13.715	&	13.645	&	$-$28.6	&	0.3	\\
16473751+4733136	&	251.906305	&	47.553795	&	13.056	&	12.394	&	12.211	&	$-$16.6	&	0.2	\\
16481312+4731205	&	252.054702	&	47.522373	&	13.987	&	13.412	&	13.294	&	0.9	&	0.3	\\
16471253+4720264	&	251.802221	&	47.340672	&	14.474	&	14.037	&	13.817	&	$-$54.9	&	2.2	\\
16470246+4726234	&	251.760290	&	47.439854	&	13.711	&	13.083	&	12.885	&	$-$13.2	&	0.6	\\
16473678+4723058	&	251.903258	&	47.384945	&	14.843	&	14.527	&	14.133	&	$-$35.8	&	0.4	\\
16472007+4729363	&	251.833642	&	47.493423	&	12.532	&	11.970	&	11.858	&	$-$162.7	&	0.2	\\
16480524+4728377	&	252.021860	&	47.477150	&	14.485	&	13.899	&	13.846	&	$-$20.1	&	0.7	\\
16464816+4750080	&	251.700676	&	47.835579	&	13.309	&	12.702	&	12.494	&	6.5	&	0.7	\\
16453637+4743149	&	251.401557	&	47.720821	&	14.597	&	14.072	&	13.854	&	$-$14.2	&	0.5	\\
16455361+4744279	&	251.473399	&	47.741108	&	14.893	&	14.324	&	14.143	&	$-$40.2	&	2.0	\\
16461661+4723453	&	251.569209	&	47.395924	&	14.464	&	13.849	&	13.715	&	$-$19.2	&	0.4	\\
16461515+4730523	&	251.563131	&	47.514542	&	14.593	&	14.060	&	13.981	&	$-$21.9	&	0.6	\\
16452323+4733276	&	251.346830	&	47.557678	&	13.025	&	12.364	&	12.212	&	$-$2.1	&	0.3	\\
16454657+4737036	&	251.444072	&	47.617680	&	14.433	&	13.785	&	13.807	&	$-$11.4	&	0.7	\\
16451881+4734262	&	251.328400	&	47.573971	&	14.547	&	13.928	&	13.840	&	$-$56.3	&	0.6	\\
16463089+4712585	&	251.628720	&	47.216267	&	13.917	&	13.328	&	13.178	&	$-$57.7	&	0.4	\\
16460964+4720273	&	251.540195	&	47.340931	&	13.574	&	12.910	&	12.685	&	$-$49.1	&	0.9	\\
16453643+4723148	&	251.401798	&	47.387466	&	14.566	&	13.918	&	13.815	&	$-$68.5	&	0.3	
\enddata

\end{deluxetable}

\clearpage
\tablenum{2}
\tablecolumns{6}
\tablewidth{0pt}

\begin{deluxetable}{cccccc}
\tablecaption{Model Atmosphere Parameters}
\tablehead{
\colhead{Star Name}     &
\colhead{T$_{\rm eff}$} &
\colhead{log(g)}      &
\colhead{[Fe/H]}      &
\colhead{$\Delta$[Fe/H]}	&
\colhead{$\xi$$_{\rm mic.}$}      \\
\colhead{(2MASS)}       &
\colhead{(K)}      &
\colhead{(cgs)}      &
\colhead{(dex)}      &
\colhead{(dex)}	&
\colhead{(km s$^{\rm -1}$)}
}

\startdata
16464310+4731033	&	4325	&	1.35	&	$-$1.08	& 0.04	&	1.70	\\
16465095+4731163	&	4000	&	0.85	&	$-$1.11	& 0.04	&	2.00	\\
16465319+4732413	&	4000	&	0.85	&	$-$1.21	& 0.04	&	2.05	\\
16465362+4730222	&	4200	&	1.15	&	$-$1.13	& 0.04	&	1.75	\\
16465422+4733371	&	3925	&	0.70	&	$-$1.04	& 0.04	&	2.00	\\
16465515+4731485	&	4025	&	0.90	&	$-$1.05	& 0.04	&	1.65	\\
16465763+4732296	&	4250	&	1.25	&	$-$1.23	& 0.04	&	1.60	\\
16470094+4731065	&	4450	&	1.60	&	$-$1.11	& 0.05	&	1.40	\\
16470210+4733163	&	4325	&	1.35	&	$-$1.19	& 0.04	&	1.80	\\
16471104+4728244	&	4225	&	1.20	&	$-$1.14	& 0.04	&	2.20	\\
16471308+4732430	&	4350	&	1.40	&	$-$1.13	& 0.04	&	1.70	\\
16472079+4731478	&	\nodata	&	\nodata	&	\nodata	& \nodata	&	\nodata	\\
16470403+4731201	&	\nodata	&	\nodata	&	\nodata	& \nodata	&	\nodata	\\
16470289+4732471	&	\nodata	&	\nodata	&	\nodata	& \nodata	&	\nodata	
\enddata

\end{deluxetable}

\clearpage
\tablenum{3}
\tablecolumns{6}
\tablewidth{0pt}

\begin{deluxetable}{cccccc}
\tabletypesize{\tiny}
\tablecaption{Line List and Adopted Reference Abundances}
\tablehead{
\colhead{Wavelength}     &
\colhead{Species} &
\colhead{E.P.}      &
\colhead{log(gf)}      &
\colhead{log $\epsilon$(X)$_{\rm \odot}$}	&
\colhead{log $\epsilon$(X)$_{\rm Arcturus}$}       \\
\colhead{(\AA)}      &
\colhead{}      &
\colhead{(eV)}      &
\colhead{}	&
\colhead{(dex)}	&
\colhead{(dex)}
}

\startdata
6300.30	&	[\ion{O}{1}]	&	0.00	&	$-$9.750	&	8.69	&	8.63	\\
6160.75	&	\ion{Na}{1}	&	2.10	&	$-$1.210	&	6.33	&	5.89	\\
6318.72	&	\ion{Mg}{1}	&	5.11	&	$-$2.010	&	7.58	&	7.38	\\
6696.02	&	\ion{Al}{1}	&	3.14	&	$-$1.520	&	6.47	&	6.28	\\
6698.67	&	\ion{Al}{1}	&	3.14	&	$-$1.910	&	6.47	&	6.28	\\
5753.62	&	\ion{Si}{1}	&	5.62	&	$-$1.104	&	7.55	&	7.38	\\
5793.07	&	\ion{Si}{1}	&	4.93	&	$-$1.974	&	7.55	&	7.38	\\
6155.13	&	\ion{Si}{1}	&	5.62	&	$-$0.764	&	7.55	&	7.38	\\
6237.32	&	\ion{Si}{1}	&	5.61	&	$-$1.075	&	7.55	&	7.38	\\
6721.85	&	\ion{Si}{1}	&	5.86	&	$-$0.956	&	7.55	&	7.38	\\
5867.56	&	\ion{Ca}{1}	&	2.93	&	$-$1.740	&	6.36	&	6.07	\\
5783.06	&	\ion{Cr}{1}	&	3.32	&	$-$0.490	&	5.67	&	5.09	\\
5787.92	&	\ion{Cr}{1}	&	3.32	&	$-$0.183	&	5.67	&	5.09	\\
5790.65	&	\ion{Cr}{1}	&	1.00	&	$-$4.053	&	5.67	&	5.09	\\
5844.60	&	\ion{Cr}{1}	&	3.01	&	$-$1.660	&	5.67	&	5.09	\\
5731.76	&	\ion{Fe}{1}	&	4.26	&	$-$1.185	&	7.52	&	7.02	\\
5732.30	&	\ion{Fe}{1}	&	4.99	&	$-$1.429	&	7.52	&	7.02	\\
5734.56	&	\ion{Fe}{1}	&	4.96	&	$-$1.744	&	7.52	&	7.02	\\
5752.03	&	\ion{Fe}{1}	&	4.55	&	$-$1.083	&	7.52	&	7.02	\\
5753.12	&	\ion{Fe}{1}	&	4.26	&	$-$0.899	&	7.52	&	7.02	\\
5759.54	&	\ion{Fe}{1}	&	4.30	&	$-$2.169	&	7.52	&	7.02	\\
5760.34	&	\ion{Fe}{1}	&	3.64	&	$-$2.514	&	7.52	&	7.02	\\
5775.08	&	\ion{Fe}{1}	&	4.22	&	$-$1.229	&	7.52	&	7.02	\\
5778.45	&	\ion{Fe}{1}	&	2.59	&	$-$3.582	&	7.52	&	7.02	\\
5793.91	&	\ion{Fe}{1}	&	4.22	&	$-$1.733	&	7.52	&	7.02	\\
5806.72	&	\ion{Fe}{1}	&	4.61	&	$-$0.946	&	7.52	&	7.02	\\
5807.78	&	\ion{Fe}{1}	&	3.29	&	$-$3.417	&	7.52	&	7.02	\\
5809.22	&	\ion{Fe}{1}	&	3.88	&	$-$1.622	&	7.52	&	7.02	\\
5811.91	&	\ion{Fe}{1}	&	4.14	&	$-$2.465	&	7.52	&	7.02	\\
5814.81	&	\ion{Fe}{1}	&	4.28	&	$-$1.904	&	7.52	&	7.02	\\
5827.88	&	\ion{Fe}{1}	&	3.28	&	$-$3.259	&	7.52	&	7.02	\\
5835.10	&	\ion{Fe}{1}	&	4.26	&	$-$2.152	&	7.52	&	7.02	\\
5837.70	&	\ion{Fe}{1}	&	4.29	&	$-$2.327	&	7.52	&	7.02	\\
5849.68	&	\ion{Fe}{1}	&	3.69	&	$-$3.046	&	7.52	&	7.02	\\
5855.08	&	\ion{Fe}{1}	&	4.61	&	$-$1.611	&	7.52	&	7.02	\\
5859.59	&	\ion{Fe}{1}	&	4.55	&	$-$0.663	&	7.52	&	7.02	\\
5862.36	&	\ion{Fe}{1}	&	4.55	&	$-$0.457	&	7.52	&	7.02	\\
5881.75	&	\ion{Fe}{1}	&	2.18	&	$-$5.177	&	7.52	&	7.02	\\
5883.82	&	\ion{Fe}{1}	&	3.96	&	$-$1.369	&	7.52	&	7.02	\\
6159.37	&	\ion{Fe}{1}	&	4.61	&	$-$1.866	&	7.52	&	7.02	\\
6180.20	&	\ion{Fe}{1}	&	2.73	&	$-$2.638	&	7.52	&	7.02	\\
6187.99	&	\ion{Fe}{1}	&	3.94	&	$-$1.736	&	7.52	&	7.02	\\
6200.31	&	\ion{Fe}{1}	&	2.61	&	$-$2.391	&	7.52	&	7.02	\\
6213.43	&	\ion{Fe}{1}	&	2.22	&	$-$2.544	&	7.52	&	7.02	\\
6221.67	&	\ion{Fe}{1}	&	0.86	&	$-$6.482	&	7.52	&	7.02	\\
6226.73	&	\ion{Fe}{1}	&	3.88	&	$-$2.199	&	7.52	&	7.02	\\
6229.23	&	\ion{Fe}{1}	&	2.85	&	$-$2.935	&	7.52	&	7.02	\\
6232.64	&	\ion{Fe}{1}	&	3.65	&	$-$1.250	&	7.52	&	7.02	\\
6246.32	&	\ion{Fe}{1}	&	3.60	&	$-$0.945	&	7.52	&	7.02	\\
6270.22	&	\ion{Fe}{1}	&	2.86	&	$-$2.674	&	7.52	&	7.02	\\
6271.28	&	\ion{Fe}{1}	&	3.33	&	$-$2.822	&	7.52	&	7.02	\\
6290.54	&	\ion{Fe}{1}	&	2.59	&	$-$4.353	&	7.52	&	7.02	\\
6290.97	&	\ion{Fe}{1}	&	4.73	&	$-$0.562	&	7.52	&	7.02	\\
6315.81	&	\ion{Fe}{1}	&	4.08	&	$-$1.709	&	7.52	&	7.02	\\
6710.32	&	\ion{Fe}{1}	&	1.49	&	$-$4.877	&	7.52	&	7.02	\\
6726.67	&	\ion{Fe}{1}	&	4.61	&	$-$1.177	&	7.52	&	7.02	\\
6733.15	&	\ion{Fe}{1}	&	4.64	&	$-$1.549	&	7.52	&	7.02	\\
6739.52	&	\ion{Fe}{1}	&	1.56	&	$-$5.030	&	7.52	&	7.02	\\
6806.84	&	\ion{Fe}{1}	&	2.73	&	$-$3.162	&	7.52	&	7.02	\\
6810.26	&	\ion{Fe}{1}	&	4.61	&	$-$1.089	&	7.52	&	7.02	\\
6828.59	&	\ion{Fe}{1}	&	4.64	&	$-$0.944	&	7.52	&	7.02	\\
6839.83	&	\ion{Fe}{1}	&	2.56	&	$-$3.364	&	7.52	&	7.02	\\
6841.34	&	\ion{Fe}{1}	&	4.61	&	$-$0.810	&	7.52	&	7.02	\\
6842.69	&	\ion{Fe}{1}	&	4.64	&	$-$1.249	&	7.52	&	7.02	\\
6843.65	&	\ion{Fe}{1}	&	4.55	&	$-$0.942	&	7.52	&	7.02	\\
6857.25	&	\ion{Fe}{1}	&	4.08	&	$-$2.228	&	7.52	&	7.02	\\
6858.15	&	\ion{Fe}{1}	&	4.61	&	$-$1.098	&	7.52	&	7.02	\\
6861.94	&	\ion{Fe}{1}	&	2.42	&	$-$3.900	&	7.52	&	7.02	\\
6862.48	&	\ion{Fe}{1}	&	4.56	&	$-$1.528	&	7.52	&	7.02	\\
5760.83	&	\ion{Ni}{1}	&	4.11	&	$-$0.760	&	6.25	&	5.81	\\
5805.21	&	\ion{Ni}{1}	&	4.17	&	$-$0.720	&	6.25	&	5.81	\\
5846.99	&	\ion{Ni}{1}	&	1.68	&	$-$3.240	&	6.25	&	5.81	\\
6175.36	&	\ion{Ni}{1}	&	4.09	&	$-$0.609	&	6.25	&	5.81	\\
6176.81	&	\ion{Ni}{1}	&	4.09	&	$-$0.260	&	6.25	&	5.81	\\
6177.24	&	\ion{Ni}{1}	&	1.83	&	$-$3.550	&	6.25	&	5.81	\\
6186.71	&	\ion{Ni}{1}	&	4.11	&	$-$0.890	&	6.25	&	5.81	\\
6223.98	&	\ion{Ni}{1}	&	4.11	&	$-$0.970	&	6.25	&	5.81	\\
6767.77	&	\ion{Ni}{1}	&	1.83	&	$-$2.100	&	6.25	&	5.81	\\
6772.31	&	\ion{Ni}{1}	&	3.66	&	$-$0.990	&	6.25	&	5.81	\\
5782.11	&	\ion{Cu}{1}	&	1.64	&	hfs	&	4.04	&	3.71	\\
5879.78	&	\ion{Zr}{1}	&	0.15	&	$-$1.350	&	2.58	&	2.08	\\
5885.62	&	\ion{Zr}{1}	&	0.07	&	$-$2.030	&	2.58	&	2.08	\\
5805.77	&	\ion{La}{2}	&	0.13	&	hfs	&	1.13	&	0.57	\\
6262.29	&	\ion{La}{2}	&	0.40	&	hfs	&	1.13	&	0.57	\\
5740.86	&	\ion{Nd}{2}	&	1.16	&	$-$0.330	&	1.42	&	0.97	\\
5811.57	&	\ion{Nd}{2}	&	0.86	&	$-$0.800	&	1.42	&	0.97	\\
5842.37	&	\ion{Nd}{2}	&	1.28	&	$-$0.401	&	1.42	&	0.97	\\
5882.79	&	\ion{Nd}{2}	&	0.56	&	$-$1.415	&	1.42	&	0.97	
\enddata

\tablecomments{For lines designated with ``hfs" we adopted the hyperfine 
structure and isotope line lists of \citet{Cunha02} and \citet{Lawler01} for 
\ion{Cu}{1} and \ion{La}{2}, respectively.}

\end{deluxetable}

\clearpage
\setlength{\hoffset}{-0.80in}
\tablenum{4}
\tablecolumns{13}
\tablewidth{0pt}

\begin{deluxetable}{ccccccccccccc}
\tabletypesize{\tiny}
\tablecaption{Chemical Abundances and Uncertainties: Oxygen to Calcium}
\tablehead{
\colhead{Star Name}	&
\colhead{[O/Fe]}      &
\colhead{$\Delta$[O/Fe]}      &
\colhead{[Na/Fe]}      &
\colhead{$\Delta$[Na/Fe]}      &
\colhead{[Mg/Fe]}      &
\colhead{$\Delta$[Mg/Fe]}      &
\colhead{[Al/Fe]}      &
\colhead{$\Delta$[Al/Fe]}      &
\colhead{[Si/Fe]}      &
\colhead{$\Delta$[Si/Fe]}      &
\colhead{[Ca/Fe]}      &
\colhead{$\Delta$[Ca/Fe]}      \\
\colhead{(2MASS)}      &
\colhead{(dex)}      &
\colhead{(dex)}      &
\colhead{(dex)}      &
\colhead{(dex)}      &
\colhead{(dex)}      &
\colhead{(dex)}      &
\colhead{(dex)}      &
\colhead{(dex)}      &
\colhead{(dex)}      &
\colhead{(dex)}      &
\colhead{(dex)}      &
\colhead{(dex)}      
}

\startdata
16464310+4731033	&	0.13	&	0.07	&	0.25	&	0.11	&	\nodata	&	\nodata	&	0.39	&	0.10	&	0.27	&	0.08	&	0.19	&	0.12	\\
16465095+4731163	&	0.63	&	0.07	&	$-$0.46	&	0.11	&	0.25	&	0.06	&	$-$0.45	&	0.10	&	0.21	&	0.08	&	0.22	&	0.12	\\
16465319+4732413	&	0.64	&	0.07	&	$-$0.58	&	0.11	&	0.24	&	0.06	&	$-$0.12	&	0.10	&	0.25	&	0.09	&	0.19	&	0.12	\\
16465362+4730222	&	0.50	&	0.07	&	$-$0.50	&	0.11	&	0.20	&	0.06	&	\nodata	&	\nodata	&	0.20	&	0.14	&	0.26	&	0.12	\\
16465422+4733371	&	$-$0.11	&	0.07	&	0.08	&	0.11	&	\nodata	&	\nodata	&	0.35	&	0.10	&	0.38	&	0.10	&	0.21	&	0.12	\\
16465515+4731485	&	0.04	&	0.07	&	0.03	&	0.11	&	0.09	&	0.06	&	0.24	&	0.10	&	0.29	&	0.12	&	0.18	&	0.12	\\
16465763+4732296	&	0.71	&	0.07	&	$-$0.33	&	0.11	&	0.34	&	0.06	&	\nodata	&	\nodata	&	0.29	&	0.08	&	0.34	&	0.12	\\
16470094+4731065	&	0.56	&	0.07	&	$-$0.15	&	0.11	&	0.15	&	0.06	&	\nodata	&	\nodata	&	0.18	&	0.09	&	0.18	&	0.12	\\
16470210+4733163	&	0.59	&	0.07	&	$-$0.30	&	0.11	&	\nodata	&	\nodata	&	\nodata	&	\nodata	&	0.18	&	0.09	&	0.33	&	0.12	\\
16471104+4728244	&	0.52	&	0.07	&	$-$0.13	&	0.11	&	0.05	&	0.06	&	\nodata	&	\nodata	&	0.27	&	0.12	&	0.06	&	0.12	\\
16471308+4732430	&	0.68	&	0.07	&	$-$0.34	&	0.11	&	0.17	&	0.06	&	\nodata	&	\nodata	&	0.20	&	0.09	&	0.19	&	0.12	\\
16472079+4731478	&	\nodata	&	\nodata	&	\nodata	&	\nodata	&	\nodata	&	\nodata	&	\nodata	&	\nodata	&	\nodata	&	\nodata	&	\nodata	&	\nodata	\\
16470403+4731201	&	\nodata	&	\nodata	&	\nodata	&	\nodata	&	\nodata	&	\nodata	&	\nodata	&	\nodata	&	\nodata	&	\nodata	&	\nodata	&	\nodata	\\
16470289+4732471	&	\nodata	&	\nodata	&	\nodata	&	\nodata	&	\nodata	&	\nodata	&	\nodata	&	\nodata	&	\nodata	&	\nodata	&	\nodata	&	\nodata	
\enddata

\end{deluxetable}

\clearpage
\setlength{\hoffset}{-0.80in}
\tablenum{5}
\tablecolumns{13}
\tablewidth{0pt}

\begin{deluxetable}{ccccccccccccc}
\tabletypesize{\tiny}
\tablecaption{Chemical Abundances and Uncertainties: Chromium to Neodymium}
\tablehead{
\colhead{Star Name}	&
\colhead{[Cr/Fe]}      &
\colhead{$\Delta$[Cr/Fe]}      &
\colhead{[Ni/Fe]}      &
\colhead{$\Delta$[Ni/Fe]}      &
\colhead{[Cu/Fe]}      &
\colhead{$\Delta$[Cu/Fe]}      &
\colhead{[Zr/Fe]}      &
\colhead{$\Delta$[Zr/Fe]}      &
\colhead{[La/Fe]}      &
\colhead{$\Delta$[La/Fe]}      &
\colhead{[Nd/Fe]}      &
\colhead{$\Delta$[Nd/Fe]}      \\
\colhead{(2MASS)}      &
\colhead{(dex)}      &
\colhead{(dex)}      &
\colhead{(dex)}      &
\colhead{(dex)}      &
\colhead{(dex)}      &
\colhead{(dex)}      &
\colhead{(dex)}      &
\colhead{(dex)}      &
\colhead{(dex)}      &
\colhead{(dex)}      &
\colhead{(dex)}      &
\colhead{(dex)} 
}

\startdata
16464310+4731033	&	$-$0.08	&	0.14	&	$-$0.13	&	0.04	&	$-$0.46	&	0.09	&	0.32	&	0.13	&	0.99	&	0.08	&	0.72	&	0.08	\\
16465095+4731163	&	$-$0.01	&	0.14	&	$-$0.06	&	0.03	&	$-$0.49	&	0.09	&	0.36	&	0.12	&	0.22	&	0.06	&	0.33	&	0.08	\\
16465319+4732413	&	$-$0.02	&	0.13	&	$-$0.08	&	0.04	&	$-$0.52	&	0.09	&	0.28	&	0.15	&	0.30	&	0.06	&	0.45	&	0.06	\\
16465362+4730222	&	$-$0.09	&	0.14	&	$-$0.09	&	0.04	&	$-$0.39	&	0.09	&	0.25	&	0.12	&	0.22	&	0.06	&	0.20	&	0.06	\\
16465422+4733371	&	0.06	&	0.13	&	$-$0.19	&	0.05	&	$-$0.48	&	0.09	&	0.87	&	0.15	&	0.85	&	0.06	&	0.86	&	0.06	\\
16465515+4731485	&	$-$0.13	&	0.14	&	$-$0.05	&	0.04	&	$-$0.45	&	0.09	&	0.21	&	0.15	&	0.25	&	0.06	&	0.26	&	0.05	\\
16465763+4732296	&	$-$0.01	&	0.14	&	$-$0.06	&	0.04	&	$-$0.49	&	0.09	&	0.40	&	0.16	&	0.48	&	0.06	&	0.39	&	0.09	\\
16470094+4731065	&	0.01	&	0.13	&	$-$0.10	&	0.06	&	$-$0.53	&	0.09	&	\nodata	&	\nodata	&	0.23	&	0.05	&	0.32	&	0.14	\\
16470210+4733163	&	0.05	&	0.12	&	$-$0.09	&	0.03	&	$-$0.46	&	0.09	&	0.27	&	0.13	&	0.36	&	0.05	&	0.42	&	0.07	\\
16471104+4728244	&	$-$0.10	&	0.14	&	$-$0.12	&	0.05	&	$-$0.52	&	0.09	&	0.43	&	0.17	&	0.26	&	0.05	&	0.36	&	0.09	\\
16471308+4732430	&	$-$0.13	&	0.13	&	$-$0.19	&	0.04	&	$-$0.52	&	0.09	&	\nodata	&	\nodata	&	0.49	&	0.06	&	0.38	&	0.06	\\
16472079+4731478	&	\nodata	&	\nodata	&	\nodata	&	\nodata	&	\nodata	&	\nodata	&	\nodata	&	\nodata	&	\nodata	&	\nodata	&	\nodata	&	\nodata	\\
16470403+4731201	&	\nodata	&	\nodata	&	\nodata	&	\nodata	&	\nodata	&	\nodata	&	\nodata	&	\nodata	&	\nodata	&	\nodata	&	\nodata	&	\nodata	\\
16470289+4732471	&	\nodata	&	\nodata	&	\nodata	&	\nodata	&	\nodata	&	\nodata	&	\nodata	&	\nodata	&	\nodata	&	\nodata	&	\nodata	&	\nodata	
\enddata

\end{deluxetable}

\end{document}